\documentclass[runningheads,a4paper]{llncs}

\usepackage{amssymb,amsmath}
\usepackage{caption}
\usepackage{graphicx}
\graphicspath{{Figures/}}  

\usepackage{gensymb}
\usepackage{url}
\usepackage[round]{natbib}
\usepackage{ upgreek }
\usepackage[dvipsnames]{xcolor}
\usepackage[framemethod=tikz]{mdframed}

\usepackage{booktabs}
\usepackage{lineno}

\usepackage{color}

\begin{document}

\mainmatter  

\title{\textit{Crossroads of the mesoscale circulation.}}

\author{Alberto Baudena$^1$\and Enrico Ser-Giacomi$^1$\and Crist{\'o}bal L{\'o}pez$^2$
\and Emilio Hern{\'a}ndez-Garc{\'i}a$^2$ \and Francesco d'Ovidio$^1$}
\authorrunning{Crossroads of the mesoscale circulation.}

\institute{$^1$Sorbonne Universit{\'e}s (UPMC, Universit{\'e} Paris 06)-CNRS-IRD-MNHN,\\
LOCEAN, 4 place JUSSIEU, F-75005 PARIS, France\\
$^2$IFISC (CSIC-UIB), Instituto de Fisica Interdisciplinar y
Sistemas Complejos, Campus Universitat de les Illes Balears,
E-07122 Palma de Mallorca, Spain }

\toctitle{Crossroads of the mesoscale circulation}
\tocauthor{Authors' Instructions}
\maketitle

\begin{abstract}
Quantifying the mechanisms of tracer dispersion in the ocean
remains a central question in oceanography, for problems
ranging from nutrient delivery to phytoplankton, to the early
detection of contaminants. Until now, most of the analysis has
been based on Lagrangian concepts of transport, often
focusing on the identification of features that minimize fluid exchange among regions,
or more recently, on network tools which focus instead on
connectivity and transport pathways. Neither of these
approaches, however, allows us to rank the geographical sites of
major water passage, and at the same time, to select them so
that they monitor waters coming from separate parts of the
ocean. These are instead key criteria when deploying an
observing network. Here we address this issue by estimating at
any point the extent of the ocean surface which transits
through it in a given time window. With such information we are
able to rank the sites with major fluxes that intercept waters
originating from different regions. We show that this
allows us to optimize an observing network,
where a set of sampling sites can be chosen for monitoring the largest flux of water dispersing out of a given region. When the analysis is performed backward in time, this method allows us to identify the major sources which
feed a target region. The method is first applied to a minimalistic model of a mesoscale eddy field, and then to realistic satellite-derived ocean currents in the Kerguelen
area. In this region we identify the optimal location of fixed stations capable of intercepting the trajectories of 43 surface drifters, along with statistics on the
temporal persistence of the stations determined in this way.
We then identify possible hotspots of micro-nutrient enrichment for the
recurrent spring phytoplanktonic bloom occuring here. Promising applications to other fields, such
as larval connectivity, marine spatial planning or contaminant detection, are then discussed.
\end{abstract}

\section{Introduction}

Suppose that a contaminant is released in a region of the open ocean. Where should a set of monitoring stations must be deployed to maximize the chances of detecting and eventually restricting the contaminant spill? Characterizing the evolution of tracers dispersed by the oceanic currents is indeed a central question in several areas of oceanography. In
some cases it is possible to partially address this point by
tracking a given tracer using satellite images or model
simulations, but often \textit{in situ} measurements and adaptive
sampling strategies are required.

Relevant examples range from the retrieval of
contaminants and their basins of attraction (\citealt{vrana2005passive,froyland2014well}),
larval dispersal and marine populations connectivity
(\citealt{carreras2017population,bradbury2008global,planes2009larval,andrello2013low,melia2016looking,monroy2017sensitivity}), the planning
of oceanographic surveys (\citealt{bellingham1996optimizing}),
monitoring systems design (\citealt{mooers2005cross,dejonge2006marine,munoz2015implication}), to the characterization of Marine Protected Areas (\citealt{cowen2006scaling,siegel2008stochatisc,shanks2009pelagic,rossi2014hydrodynamic,dubois2016linking,bray201773}), or to the so called Marine Spatial Planning (MSP, \citealt{viikmae2011spatial,delpecheellmann2013investigating,ehler2018marine}) which integrates all the above.

Horizontal transport at the ocean surface is one of the key
mechanisms controlling the dispersion of tracers, in particular on scales of the order of days to weeks. In this time frame, a heuristic but very common assumption is that vertical velocities are weak enough so that the motion of a water parcel can be considered two-dimensional. Strictly speaking, horizontal transport affects any advected tracer through two main processes: mixing, which reduces and smooths the gradients, and
horizontal stirring, which instead enhances the tracer gradients
(\citealt{eckart1948analysis,okubo1978horizontal,garrett1983initial,sundermeyer1998lateral}).
Horizontal stirring is one of the main mechanism
for which an initial homogeneous water mass is stretched by the
currents and deformed into elongated and convoluted
filaments. These features can
intrude into some regions far apart from their origin, while
other areas close to the source location may be shielded by
circulation features, the so-called transport barriers.
This filamentation process eventually enhances
mixing by creating longer contact surfaces between
water masses of contrasting properties.

Several tools have been proposed to achieve a better
understanding of the role horizontal transport plays in tracer dispersion.
Most of these approaches focus on the Lagrangian analysis
of the ocean currents based on dynamical system theory
(\citealt{ottino1989kinematics,haller_lagrangian_2000,mancho2004computation,shadden2005definition,wiggins2005dynamical}).
These methods are mainly devoted to the detection of barriers to transport in flow systems
(\citealt{boffetta_detecting_2001,abraham_chaotic_2002,dovidio_mixing_2004,beronvera2008oceanic,prants2014identifying,prants2014lagrangian})
or coherent regions
(\citealt{lehahn_stirring_2007,froyland2007detection,d2013ecological,
berline2014connectivity,hadjighasem2016spectral,miron2017lagrangian})
with minimal leaking to surrounding water masses. Interestingly, during the last years, network theory approaches to geophysics (\citealt{philipps2015graph}), flow transport and mixing (\citealt{ser2015flow,lindner2017spatio,fujiwara2017perturbation,molkenthin2017edge,padberg2017network,ser2017lagrangian,mcadam2018surface}), and turbulence (\citealt{iacobello2018visibility,meena2018network}) have also attracted lot of interest.

Recently, more attention has been given to a
problem that is complementary and somehow opposite to the identification of transport barriers: how to detect regions that enhance fluid exchanges
across a flow system (\citealt{ser2015most,ser2015dominant,costa2017calculation,lindner2017spatio,rodriguez2017clustering,koltai2018large})? This issue has
been developed particularly in the framework of networks built from Lagrangian trajectories (\citealt{ser2015flow}). In such studies, spatial
sub-areas of the ocean are represented by network nodes while
links parameterize the amount of water exchanged among
them. Networks constructed in this way, called Lagrangian Flow
Networks (LFNs), provide thus a topological description of the
transport dynamics in the ocean allowing to describe features of the flow by using several network methods and tools.

In particular, it has been shown the existence of localized
\textquotedblleft bottlenecks\textquotedblright whose
presence maintains connections between
areas of the seascape that would be otherwise almost
disconnected. Borrowing from Network Theory the concept of
betweenness centrality, it has been possible to quantify
explicitly how much LFN nodes act as bottlenecks for the
oceanic flow by counting the number of paths passing through
them during a fixed interval of time (\citealt{ser2015most}).

The main limitation of the betweenness centrality concept, however, is
that it identifies bottleneck regions that, even if
topologically relevant for the connectivity geometry, do not
present necessarily an important water flux. This is
related to the fact that the betweenness measures
the number of connections crossing a node, regardless of connectivity
strength. Therefore, to pinpoint hotspots that maximize the
incoming or outgoing flux in a given time window, one should
focus on strong \textquotedblleft passage
points\textquotedblright of water, instead of only topological
bottlenecks.

In this paper we aim to detect these hotspots, which
act as \textquotedblleft crossroads\textquotedblright of the
circulation, being traversed by the largest amount of water coming from
(or going to) a focal region. This is achieved by defining a quantity, which we call
\textit{crossroadness}, which is the surface flow through each
spot of the domain considered in a specific temporal window.

Moreover, our calculation
preserves the information on the original location of each passing
trajectory. Based on the \textit{crossroadness} analysis, we introduce an algorithm for designing a
sampling network with minimal redundancy, i.e. in which the
flow of a given source region through the network is
maximized and the number of stations minimized. Reversing the analysis backward in time, we can use the same
method to find the major \textquotedblleft
source\textquotedblright points from which the water
distributes over a target area.
The paper is organized as follows: Sec. \ref{sec:data}
describes the methodological framework, the theoretical model and the dataset
employed for the computation and the validation of the results.
In particular, Subsec. \ref{subsec:CR} introduces the
crossroadness and its fluid dynamical and oceanographic interpretation. We then determine a
ranking method (Subsec. \ref{subsec:CRstations}) that provides
the sites with the major passage of water dispersed from a
source region or feeding a target region, when computed forward
or backward in time, respectively. Validation and case studies
are then described in Sec. \ref{sec:results}. In Subsec. \ref{subsec:NS2Dres},
we first analyse the crossroadness properties of a simple steady vortex configuration obtained from the 2D Navier Stokes equation.
In Subsec. \ref{subsec:svp}, the method is applied to satellite altimetry
data and validated against the trajectories of real SVP
drifters in the Southern Ocean. In Subsec.
\ref{subsec:sur_vs_st}, the relationship between the surface
being monitored and the number of stations employed is examined. An
analysis of the persistence of the observing network is reported in
Subsec. \ref{subsec:persistence}. In Subsec.
\ref{subsec:source_region}, we use the method to identify
possible sites from which nutrients are delivered offshore the
Kerguelen plateau, fuelling the open ocean planktonic bloom. A
summary of the results, along with an illustration of the
perspectives and possible fields of application is given in the
Discussion (Sec. \ref{sec:discussion}). Finally, our paper is
completed by two appendices, extending the analysis of the
sensitivity of our diagnostic to changes in the dates of the
velocity field used (Appendix A), and giving some demonstration of the link of our
new diagnostic, under particular hypotheses, to averages of
velocity and of kinetic energy (Appendix B).
\begin{figure}[t]
\centering
\includegraphics[width=12.0cm]{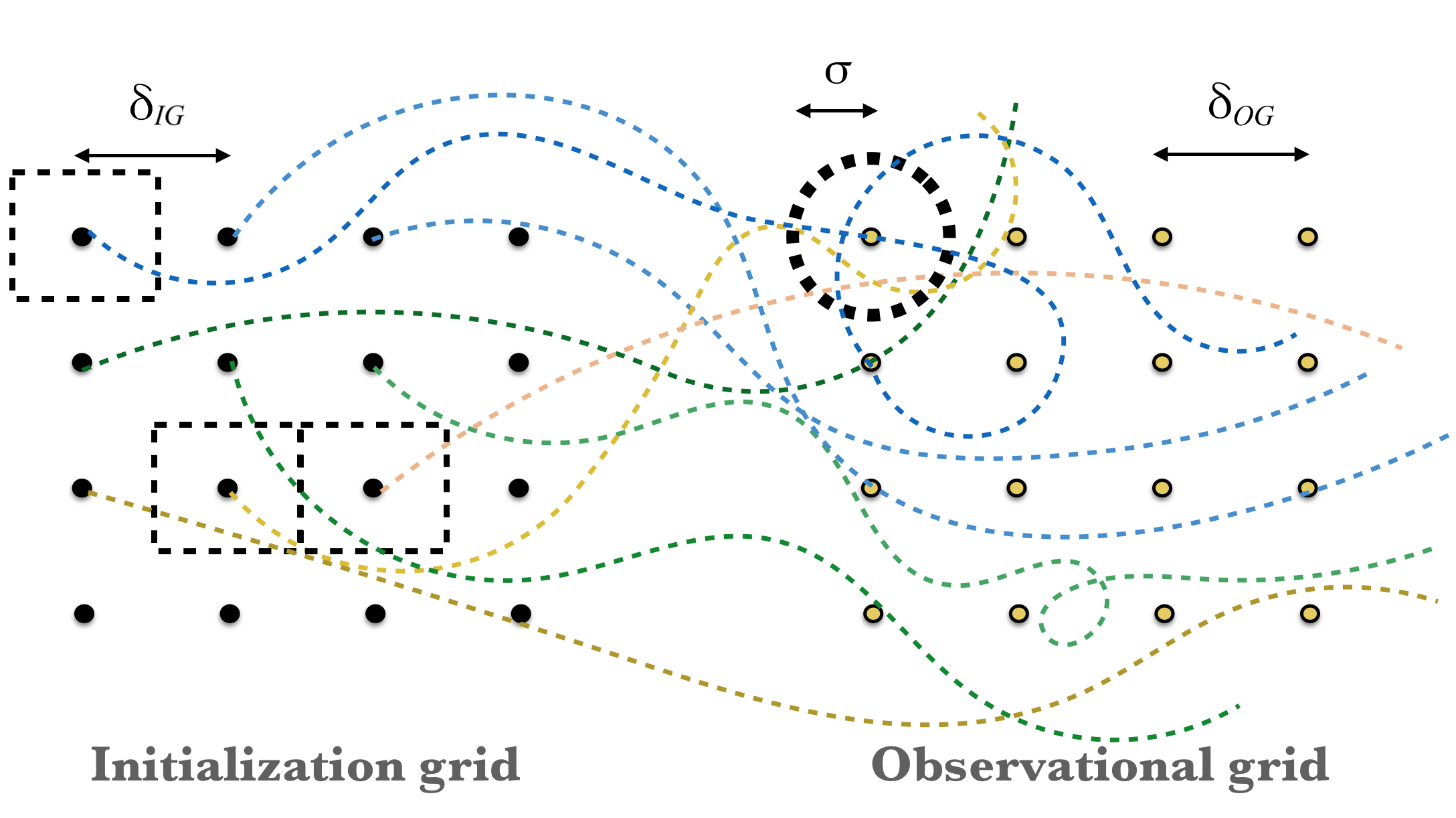}
\internallinenumbers
\caption{Illustrative example for the calculation of the
crossroadness. Colored dotted lines correspond to different
trajectories originating from the \textit{initialization grid} (left),
advected for a time $\uptau$. The circle (thick dotted line)
represents the detection range of the station of the \textit{observational grid}
(right), while the rectangles (thin dotted line) contain the surface of water
that passes through the station.}
\label{fig:scheme_CR}
\end{figure}

\section{Data and Methods}
\label{sec:data}

\subsection{The crossroadness: characterizing regions by the amount of water crossing them}
\label{subsec:CR}

We define a new metric, that we call
\textit{crossroadness} (CR), measuring the surface
crossing a region of a given size in a fixed temporal window.

To do this, we consider two grids, the \textit{initialization}
and the \textit{observational grid} {(Fig.
\ref{fig:scheme_CR})}, of cell sizes $\delta_{IG}$ and
$\delta_{OG}$ respectively, where each cell is represented by
its central point. If $\delta_{IG}$ is small enough, we can
consider every central point as representative of the content
of the cell.

Each central point of the \textit{initialization grid} is advected and defines a trajectory. The \textit{observational grid} represents instead the domain over which we compute the crossroadness (see also Subsec. \ref{subsec:grids}). For each point $\mathcal{P}$ of this grid, we count the number of trajectories passing below a distance $\sigma$ from $\mathcal{P}$. This is done by
computing the Euclidian (or spherical when working with spherical coordinates) distance between $\mathcal{P}$ and the
first point of the trajectory, then the second one, the third
one and so on. We obtain thus $N_{PT}$ values. We consider as distance
between the trajectory and $\mathcal{P}$ the
minimum among the $N_{PT}$ values. $\sigma$ represents the neighborhood radius (the
detection range) of each point in the \textit{observational grid}. In spherical coordinates, if $\sigma$ is given in radians, this radius is indeed $R \sigma$. As
the dimensions of a cell in the \textit{initialization grid}
are relatively small, we can consider that the trajectory is
representative for the whole content of the cell from which it
is originated (therefore, $\sigma\geq\delta_{IG}$). Thus, if we
multiply the number of trajectories counted times the surface
of an initialization cell, $\Delta$, we will obtain an estimation of the total
water surface that passed through the node during the period
$\uptau$. We call this total water surface flowing inside the
detection range of the point $\mathcal{P}$ the
\textquotedblleft crossroadness\textquotedblright
(\textquotedblleft CR\textquotedblright) at $\mathcal{P}$. In
this way we define a CR field on all points of the
\textit{observational grid}. A
representative scheme of this concept is illustrated in Fig.
\ref{fig:scheme_CR}, where the crossroadness of the point in
the circle is $3\Delta$ since there are three trajectories
entering its detection range. Because of the quasi-twodimensional nature
of ocean circulation at the scales considered here, the amount
of surface is proportional to the volume and then to the mass
of the water transported in the upper ocean layers.
The same procedure can be applied backwards: the
\textit{initialization grid} is advected backward in time, and,
over the \textit{observational grid}, we count for each point
how many trajectories pass closer than the detection range $\sigma$.
In this way we obtain the crossroadness field backwards in time.

\subsection{Theoretical relation of the crossroadness with absolute velocity and mean kinetic
energy}
\label{subsec:CR_Uabs_Ek}

Intuitively, one can expect higher crossroadness values in points where the velocity is higher, because the corresponding particle fluxes are also larger. The situation is in fact more complex because a map of crossroadness is associated only to the trajectories stemming from a specific region. It is however instructive to study the case in which the trajectories stem from the entire domain (i.e., when the initialization grid coincides with the observational grid), because in this case the crossroadness has a direct proportionality relation with the kinetic energy.

To show this relation, we note that every circle of radius $\sigma$
around an observational station presents a cross section (more
properly a cross-length) $2\sigma$ perpendicular to the flow
coming from any direction. If $\sigma$ is sufficiently small to
allow considering the velocity field constant on the whole
observational circle, the amount of surface crossing that
station in a short interval of time $dt$ is $2\sigma |v|dt$,
with $|v|$ the velocity modulus at the station. Integrating
during a time $\uptau$ we find that the CR at that point can be
written as
\begin{equation}
   CR=\int_{D_0}^{D_0+\uptau} 2\sigma |v| dt =2\sigma\uptau <|v|> \ ,
\label{eq_CR_v}
\end{equation}
with $<|v|>$ the temporal average of the speed $|v|$ in the
time interval $\uptau$.
If the temporal fluctuations of
$|v|$ during the time interval $\uptau$ can be neglected, i.e.
$<v^2> \approx <|v|>^2$ then a relationship with the temporal
average of the kinetic energy per unit of mass of the flow,
$<E_K>=<v^2/2>$, would hold:
\begin{equation}
CR \approx  2\sigma \uptau \sqrt{2<E_K>}.
\label{eq_CR_EK_2}
\end{equation}
The formulas above present the announced
approximate relationship between the crossroadness and the flow
speed. For a validation of Eq. (\ref{eq_CR_v}) and (\ref{eq_CR_EK_2}), please refer to Appendix B.

As we shall see, more complex crossroadness patterns arise when the trajectories do not originated from the entire domain.

\subsection{Ranking method for the optimization of a network of CR stations}
\label{subsec:CRstations}

\begin{figure}[b!]
\centering
\includegraphics[width=3.875cm]{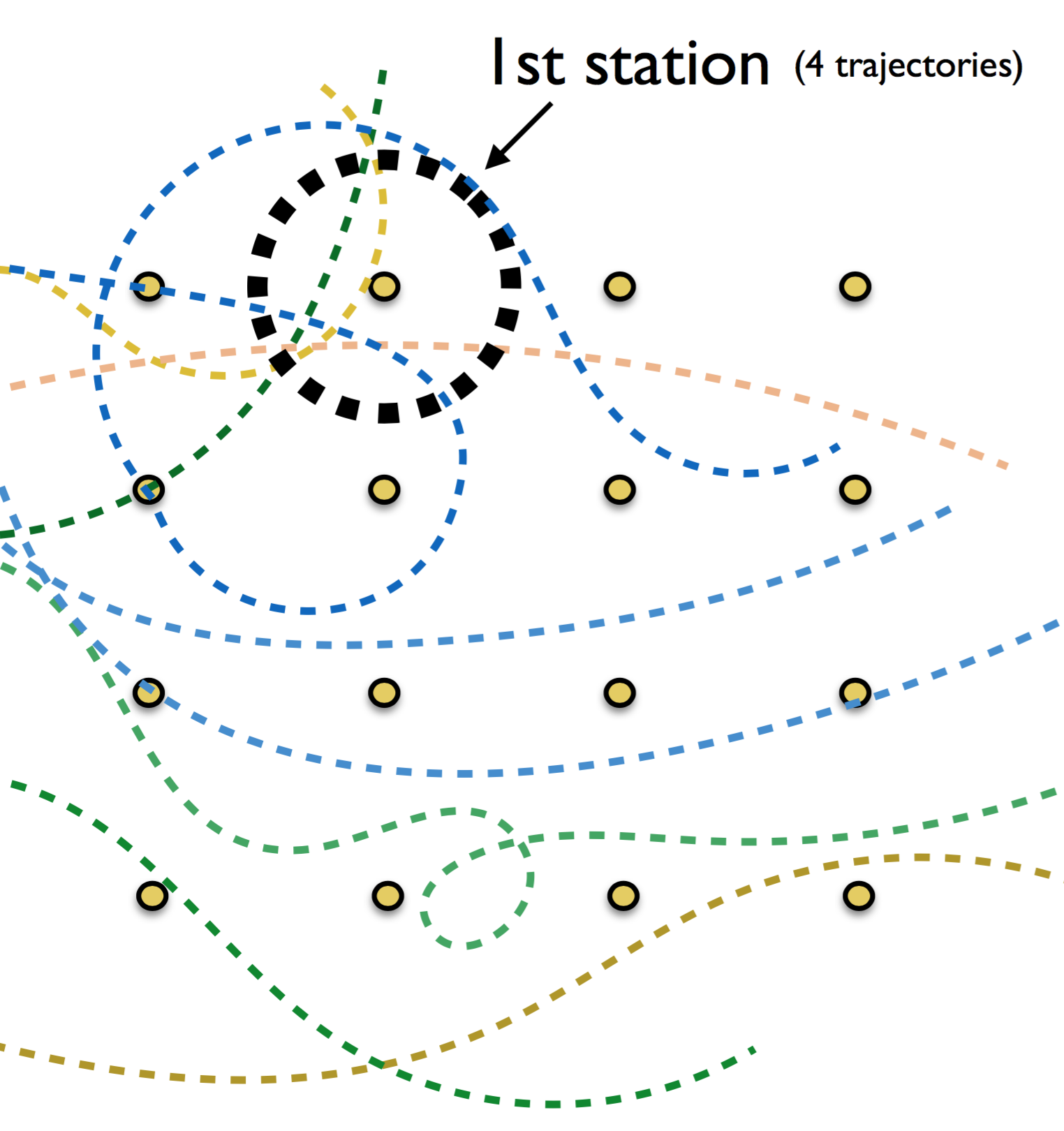}\quad\includegraphics[width=3.875cm]{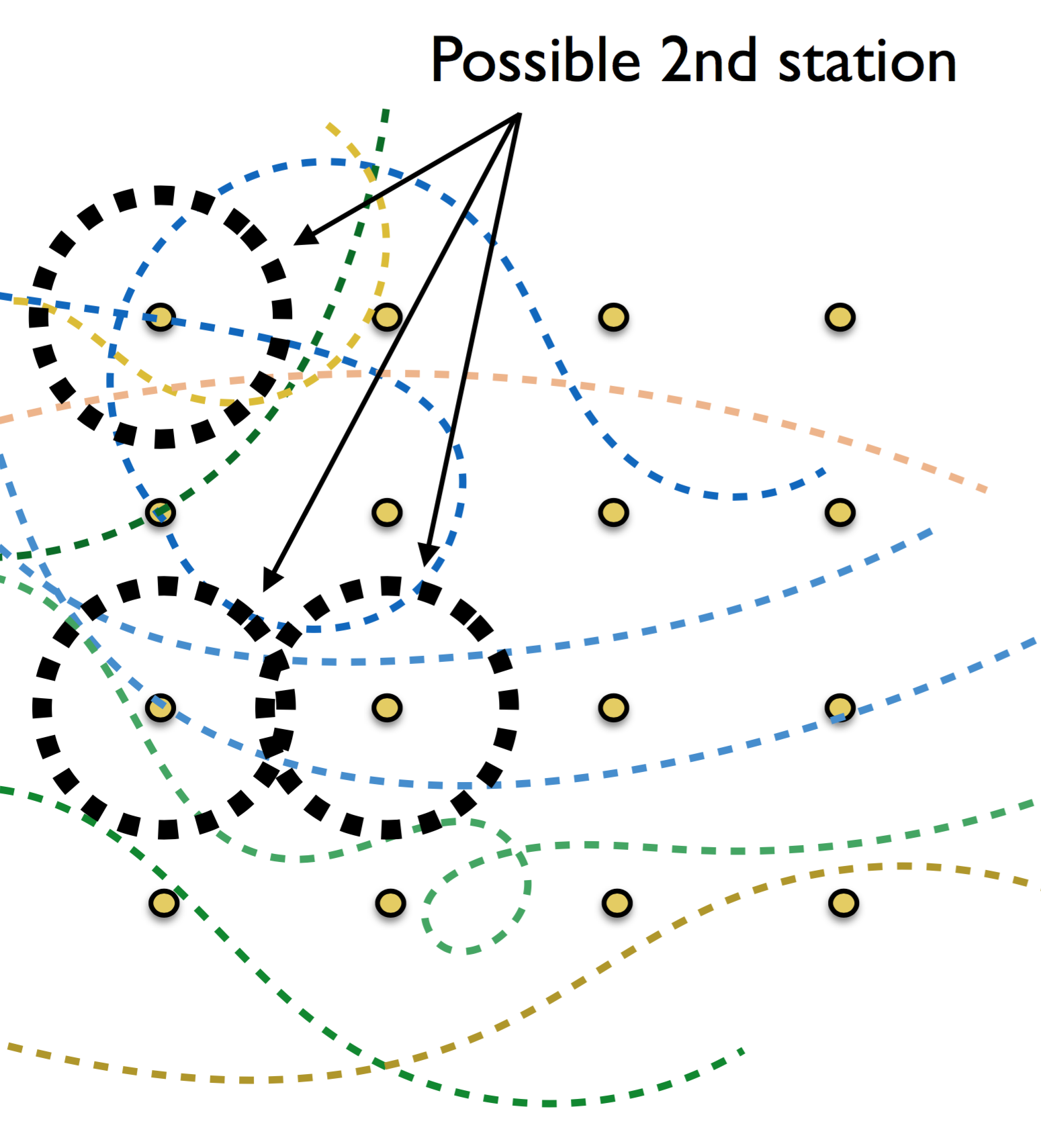}\includegraphics[width=3.875cm]{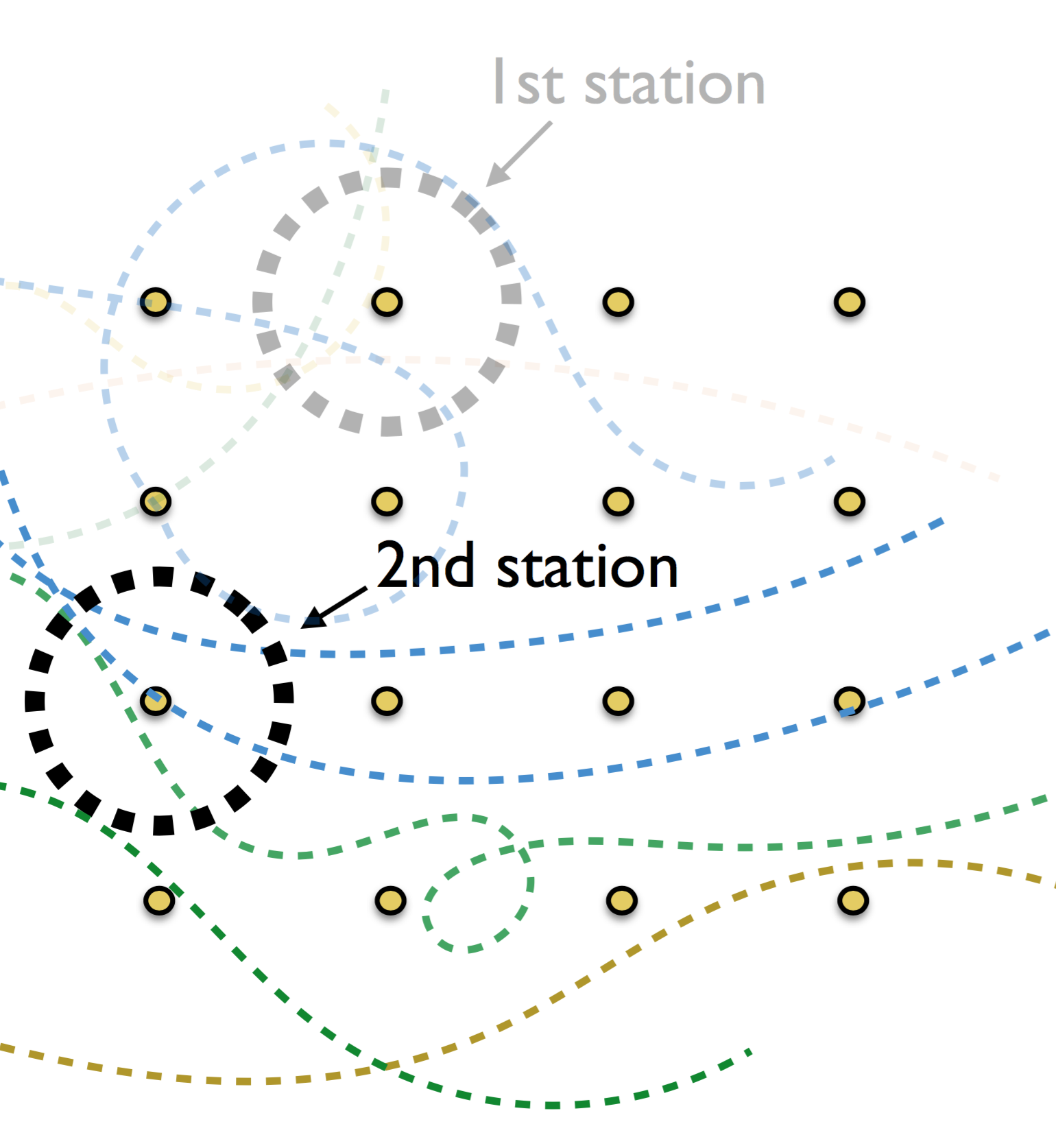}
\internallinenumbers
\caption{Illustrative scheme for the determination of the position of a
network of observing stations which maximize the detection of a dispersed tracer.
The first station is the one that collects the largest number of trajectories,
in this case the one circled in the left panel (4 trajectories). Then, we have 3 possible
second stations, each with 3 trajectories passing near them (middle panel). But since we
consider only independent trajectories, we have to exclude the ones already sampled by
the first station (panel on the right). In this way we have only one second possible station.}
\label{fig:Stations_scheme}
\end{figure}

In the study of the dispersion of a passive tracer one of the
main questions is the definition of an effective sampling
strategy. Considering the tracer (e.g. a contaminant) with a
finite life time, we want to know what would be the best
distribution of monitoring sites (referred to as \textquotedblleft
stations\textquotedblright) that can intercept the largest
fraction of the advected tracer. In a turbulent pattern of
circulation, the answer to this question is not obvious, since
the patch can be redistributed irregularly through the domain
considered.\\
Sampling on a regular grid is a possible choice. However,
depending on the circulation patterns, we can easily imagine
that some retrieving sites may convey water from larger regions
than others. The crossroadness provides
implicitly a simple method for the definition of a sampling
strategy. In fact, if we want to choose the best monitoring
station, we will simply select the site crossed by the largest
number of trajectories originated by the source region of the
tracer, i.e. the one with the higher value of crossroadness. In
order to define the second most important monitoring station,
we exclude the trajectories already monitored by the first
station. We consider only independent trajectories, i.e.
originated from different locations of the
\textit{initialization grid}, and we determine also the second
station. Proceeding iteratively in this way we identify a network of
\textit{CR stations} (Fig. \ref{fig:Stations_scheme}).\\
The calculation can be performed backward in time as
well. The initialization region becomes a target
region. In this case, the network of CR stations represents
the ensemble of locations which maximize the surface water
feeding the target region.

\subsection{A simple steady vortex field}
\label{subsec:NS2D}

We consider a numerical simulation of the Navier Stokes equation in two dimensions, which, for the scalar vorticity field  $\omega=\nabla \times \textbf{u}$ of an incompressible fluid ($\nabla \cdot \textbf{u}=0$, in which \textbf{u}=(u,v) is the velocity field), can be written as:
\begin{equation}\label{eqNS2D}
\frac{\partial \omega}{\partial t} + \omega\cdot \nabla \textbf{u}\,=\,\nu\nabla^2 \omega -\alpha\omega +f_{\omega}
\end{equation}
in which $\nu$ is the kinematic viscosity, $-\alpha \omega$ is a friction term removing energy at large scales and $f_{\omega}$ represents a forcing term necessary to maintain a stationary state. All the quantities are adimensionalised. We use a forcing on a narrow band of wave numbers around $k_f$=2 in Fourier space which is constant in time and acting with a fixed phase $\theta=\pi$ . We take $\nu=0.5$, $\alpha=0.025$ and we integrate Eq. \ref{eqNS2D} with a pseudospectral code on a box of size $L_x=L_y=2\pi$ with periodic boundary conditions and 512$^2$ grid points. Starting from a null vorticity field,  the model is run for 15600 steps (dt$=5\cdot10^{-4}$, t$\,=7.8$), until the energy of the system is constant ($E=0.06$) and a steady  vorticity (and velocity) field is reached. More details on the model are provided in  \citealt{boffetta2002intermittency,boffetta2010evidence}.

\begin{figure}[h!]
\centering
\includegraphics[width=12.0cm]{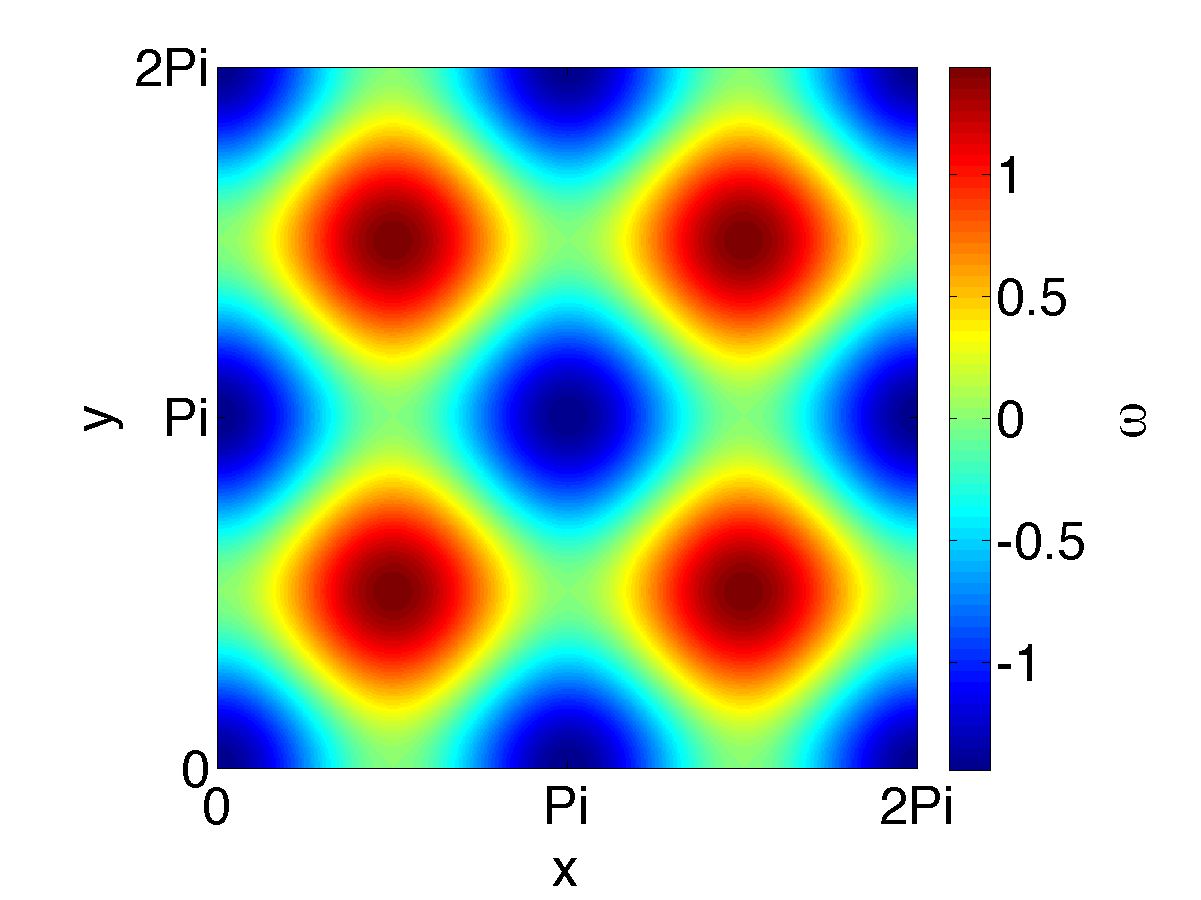}
\internallinenumbers
\caption{Two-dimensional steady vorticity field as obtained from Eq. \ref{eqNS2D}. The box size are $L_x=L_y=2\pi$ and the vorticity field was integrated with a resolution of 512$^2$ grid points regularly spaced.}
\label{fig:vortNS2D}
\end{figure}

The stationary vorticity field obtained is  in Fig. \ref{fig:vortNS2D}. It is formed by a series of anti-clockwise (in blue) and clockwise (in red) rotating vortexes. Particles, disposed on a regular grid of 256$^2$ points, are then advected using a second order Runge Kutta method as explained in Subsec \ref{subsec:grids}. The vorticity field is considered to cover periodically the full horizontal plane, on which the particle move, although we will only display and analyze the subdomain [0,2$\pi$]$\times$[0,2$\pi$]. Finite Time Lyapunov Exponents (FTLE) forward ($\lambda^f$) and backward ($\lambda^b$) in time are computed from the trajectories, following the definition in \citet{shadden2005definition}.

\subsection{Ocean satellite-derived velocity field and chlorophyl data}
We focus on the Kerguelen region of
the Southern Ocean, where in 2011 the KEOPS2 campaign
released, over 20 days, a set of 48
Surface Velocity Program (SVP) drifters. We use a set of stations, defined by our
method, for a virtual \textquotedblleft search and
rescue\textquotedblright of these drifters. For their computation, we use satellite-derived velocity fields, which are obtained through the geostrophic approximation from the measurement of the Sea Surface Height (SSH).
This approximation and the relatively low spatial and temporal resolution of satellite products
often risk to smooth mesoscale structures (\citealt{bouffard2014lagrangian}), especially close to the coast (\citealt{nencioli2011surface}).
The comparison between satellite-derived crossroadness and real drifter trajectories is thus a realistic assessment of the capacity of the algorithm
described here to identify points with enhanced passage of
particle trajectories.

The Kerguelen region is also
characterized by a massive spring-time phytoplanktonic bloom, extended for hundreds of kilometres,
preconditioned by the redistribution of micro-nutrients (in
particular iron) advected by the Antarctic Circumpolar Current (ACC)
from the Kerguelen plateau margin out into the open ocean
(\citealt{d2015biogeochemical}). Satellite-derived chlorophyll
is therefore a useful tracer of transport pathways, and an
opportunity for pinpointing likely sources of iron by employing
the theoretical concepts developed in this paper.

For the velocity field we use the DUACS (Data Unification and
Altimeter Combination System) delayed-time multi-mission
altimeter gridded products defined over the global ocean with a
regular $\frac{1}{4}\times\frac{1}{4}\degree$ spatial sampling
(\citealt{pujol2016duacs}) and available from the E.U.
Copernicus Marine Environment Monitoring Service (CMEMS,
http://marine.copernicus.eu/). An experimental product, which used an improved mean dynamic topography and for which the altimetric tracks have been interpolated with optimized parameters,
is also considered. It corresponds to the AVISO regional Kerguelen
product, velocities computed from altimetry in delayed time on
a (higher resolution) $\frac{1}{8}\times\frac{1}{8}\degree$ regular grid
(http://www.aviso.altimetry.fr/duacs/).

The field of chlorophyll was downloaded from the Oceancolour
product
(OCEANCOLOUR\_GLO\_CHL\_L4\_REP\_OBSERVATIONS\_009\_082)
at CMEMS. These data are provided by a map computed from satellite
observations over a period of 8 days (to limit cloud coverage),
and with a spatial resolution of $4$ km$\times\,4$km.\subsection{SVP drifters}
During the KEOPS2 campaign in October-November 2011, 48 Surface
Velocity Program (SVP) drifters were released within the Kerguelen
region. The SVP drifting buoy is a Lagrangian
current-following drifter, composed of a spherical surface
float of 35 cm in diameter, which contains the battery, a holey sock
drogue of about 6 meters
that tracked the water currents at a nominal depth of 15 m,
representative of the surface circulation, and a satellite
transmitter which relays the data through the Iridium system.
All the buoys address the World Ocean Circulation
Experiment (WOCE) benchmarks.

In our study we consider only the drifters released on the eastern part
of the Kerguelen plateau, i.e. at a longitude greater than $68\degree$E. We therefore consider 43 drifters
trajectories of the 48 drifting buoys originally deployed. We assume the 11th of
November 2011 as the release date. The duration of the trajectories
ranges between 63 to 93 days (average of 81 days), with a
temporal resolution of 1 day.

\subsection{Initialization and observational grids}
\label{subsec:grids}

We describe here the details of the construction of the two grids used for the computation of the crossroadness.

For the Navier Stokes model, particles start over an \textit{initialization grid} of 256$^{2}$ regularly displaced points ($\delta_{IG}=0.0246$) and are advected with a second order Runge-Kutta scheme for a period t=11.7, corresponding to 23400 steps.

When working with altimetric velocities, the points of the \textit{initialization grid} are  advected using DUACS altimetric velocities with a 4th order Runge-Kutta integration
scheme, for a period between 30 and 90 days and a time step of
3 hours. Each trajectory thus computed contains 8 points per
day ($N_{PT}\,=\,8\,N_D$ in total, $N_D$ number of days). We remind that the period of integration is chosen so that the vertical motion can be neglected. Indeed, below 2-3 months, and excluding some peculiar regions of convection, the vertical velocity is usually 2$\sim$3 orders of magnitude smaller than the horizontal one, and thus the ocean surface motion can be approximated as two dimensional (\citealt{dovidio_mixing_2004,rossi2014hydrodynamic}).

We note here that, when working with altimetric velocities, and thus in spherical coordinates, we impose that all the cells have the same size. The initial angular separation between two contiguous particles
($\delta_{IG}$) used in the computations varies from 0.1 to 0.4 degrees. By initial separation
$\delta_{IG}$ we mean the difference in latitude among contiguous rows of
particles. Along the longitudinal
(LON) direction, to keep the same distance, the separation is thus corrected
by the cosine of the latitude LAT
($\delta_{LON}=\delta_{IG}/\cos(LAT)$). In this way all the cells
have the same lateral sizes, $R\delta_{IG}$, and area
$\Delta=R^2\delta_{IG}^2$ (with $\delta_{IG}$ in radians, and $R$ the
Earth radius).
The \textit{observational grid} in spherical coordinates is built in the same  way as the \textit{initialization grid}.

In Sec. \ref{sec:results} we will always refer to them as \textit{initialization} or \textit{observational grid}, but keeping in mind their difference (for Navier Stokes and altimetric velocities) in Euclidian and spherical coordinates.

\section{Results}
\label{sec:results}

\subsection{The crossroadness in the steady Navier-Stokes 2D vortex configuration}
\label{subsec:NS2Dres}

\begin{figure}[t!]
\centering
\includegraphics[height=4.4cm]{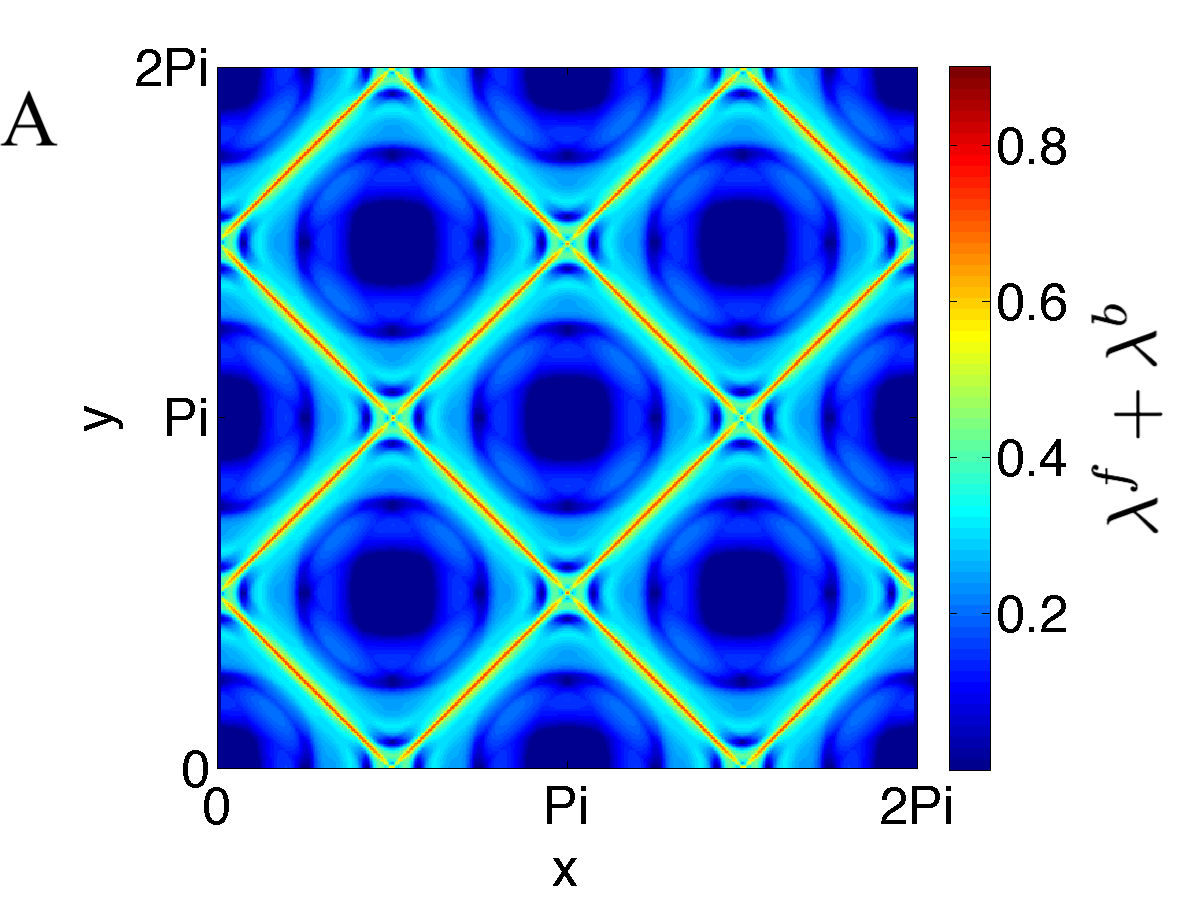}
\quad
\includegraphics[height=4.4cm]{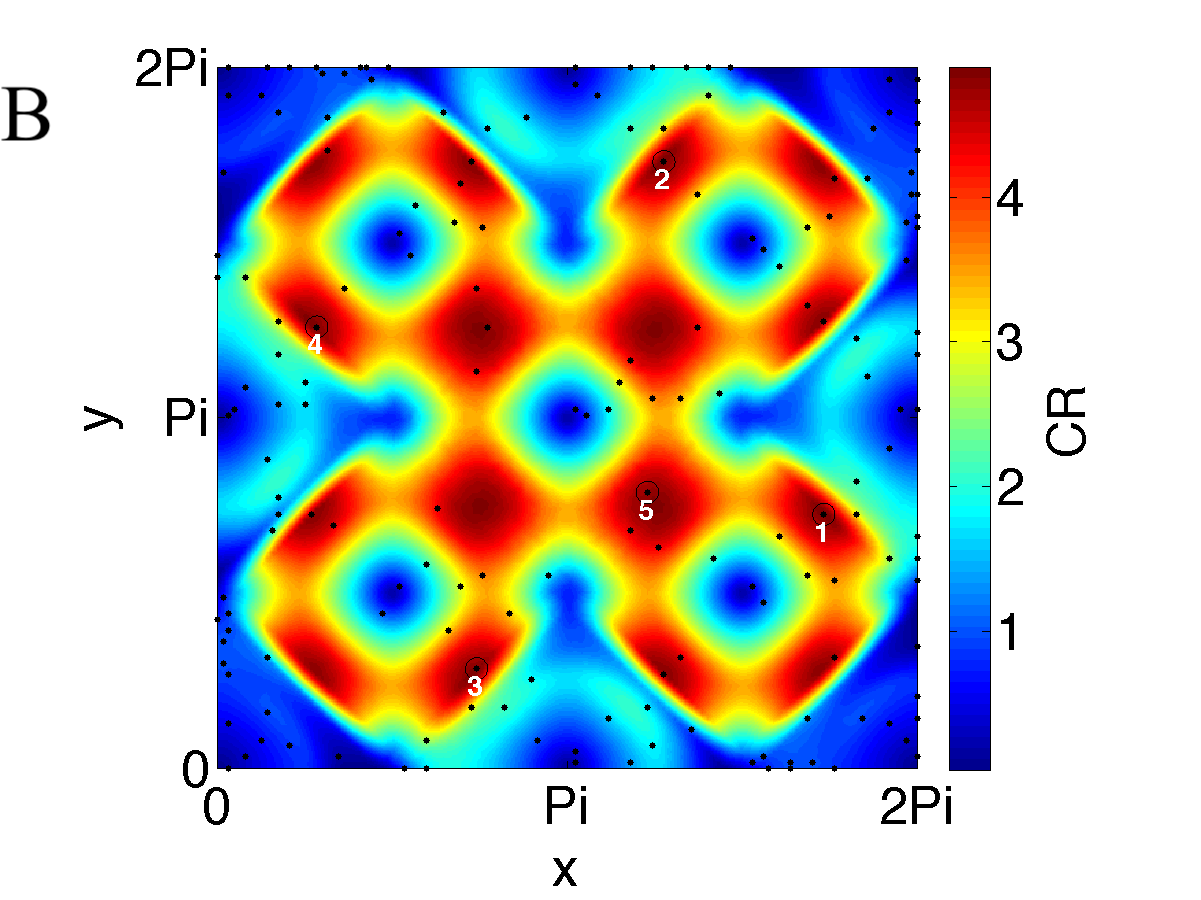}
\quad
\includegraphics[height=4.4cm]{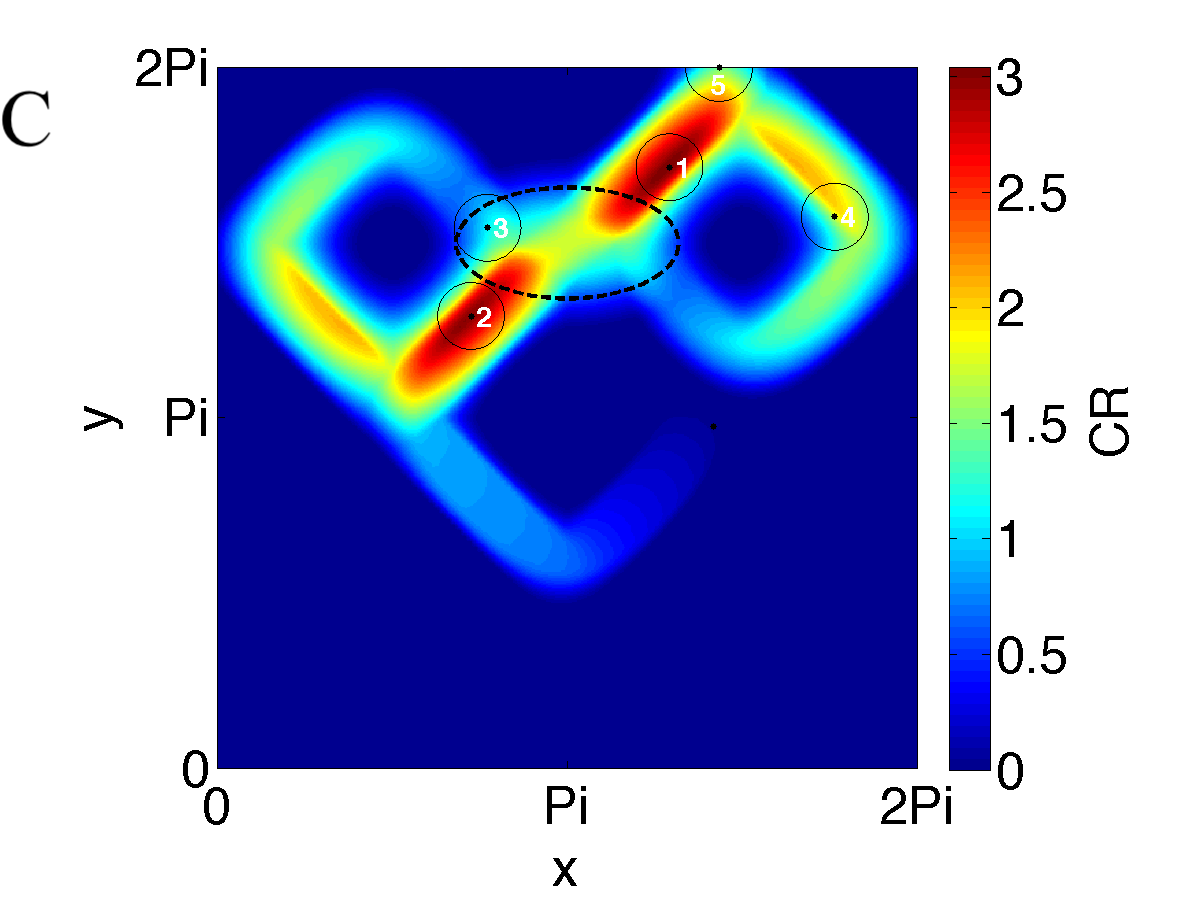}
\quad
\includegraphics[height=4.4cm]{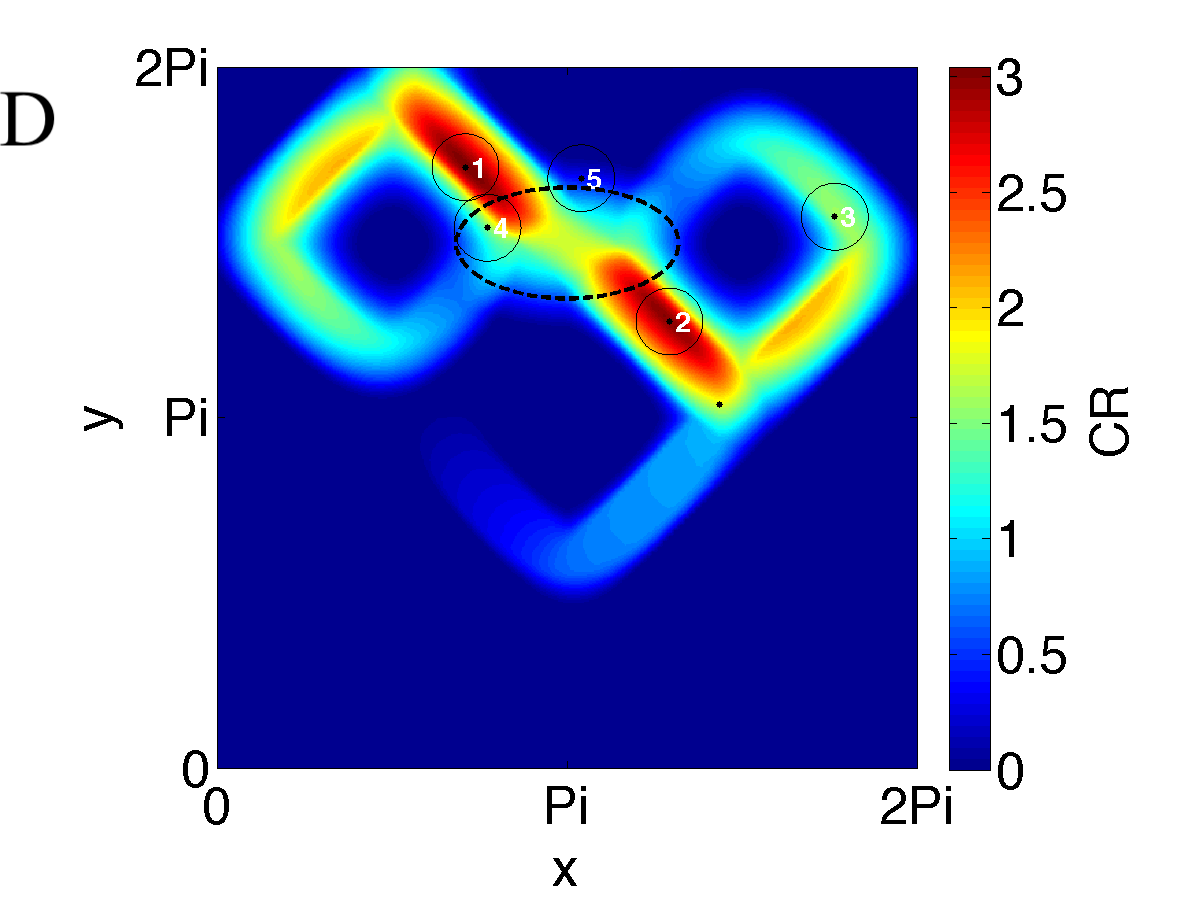}
\internallinenumbers
\caption{Panel A: Forward ($\lambda^f$) and backwards ($\lambda^b$) FTLE fields, computed with integration times $t=\pm11.7$ respectively, and over a 256$^2$ regular grid. They identify the locations of stable and unstable manifolds and the associated hyperbolic points. The quantity actually plotted is $\lambda^f+\lambda^b$ Panel B: crossroadness forward in time, obtained from the advection of a 256$^2$ regular \textit{initialization grid} {(covering the whole domain)} for a time $t=11.7$, computed over a 128$^2$ \textit{observational grid} covering the same domain. $\sigma=0.1$. Panel C-D: crossroadness, forward and backward in time respectively, computed from the advection of the points started at the ellipse (black dotted line) centered in $x_C=\pi$, $y_C=\frac{3}{2}\pi$, with semi-axes $r_x=1$, $r_y=0.5$ and spatial separation between two contiguous particles $\delta_{IG}=0.0246$, for a time $t=\pm11.7$, respectively. The \textit{observational grid} is the same of panel B. For B,C,D, each black dot represents a CR station. The detection range of each of the 5 most important stations is displayed as a black circle, and contains the order of priority of each station (white number).}
\label{fig:NS2Dres}
\end{figure}

To gain some insights on the interpretation of the crossroadness and on its differences in respect to other Lagrangian methods, we present here the analysis of the flow associated to the vorticity field shown in Fig. \ref{fig:vortNS2D}, obtained from the numerical simulation of Eq. (\ref{eqNS2D}), as explained in Subsec. \ref{subsec:NS2D}. The flow contains 4 vortices rotating clockwise (positive vorticity, red patches) and a central vortex rotating anti clockwise (in blue). 9 other anti clockwise rotating vortices are present partially on the edges of the box. The vortices are surrounded by separatrices made of the stable and unstable manifolds of fixed hyperbolic points, arranged in a square lattice. To visualize these separatrices we compute in Fig. \ref{fig:NS2Dres} Finite Time Lyapunov Exponent fields (FTLE, defined as in \citealt{shadden2005definition}). Forward ($\lambda^f$) and backwards ($\lambda^b$) FTLE are calculated for an integration time $\tau=\pm11.7$ on a grid of 256$^2$ points, and we plot their sum. The ridges of such field clearly identify the stable and unstable manifolds and its maxima highlight the position of the correspondent hyperbolic points.

As a first step, we compute the crossroadness field and the associated monitoring stations considering as \textit{initialization grid} 256$^2$ particles regularly displaced on the entire domain ($\delta_{IG}=0.0246$). The \textit{observational grid} is made of 128$^2$ points ($\delta_{OG}=0.0490$). Particles in the initialization grid are advected forward in time for t$=11.7$, i.e. the same period considered for the computation of the FTLE of Fig. \ref{fig:NS2Dres}A. Finally, we consider as detection range $\sigma=0.1$. The result is shown in Fig. \ref{fig:NS2Dres}B. The crossroadness pattern in itself does not contain much valuable information, because (as explained in Subsec. \ref{subsec:CR_Uabs_Ek}) when the \textit{initialization grid} covers the entire domain the crossroadness closely matches the kinetic energy (except for points very close to the border). The position of the monitoring stations selected by the ranking algorithm (the set of points crossed by the largest number of independent trajectories) however is not trivial. They are shown as black dots in each panel. For the first five, we plotted also their detection range and order of importance (white number inside the circle). Although in general FTLE ridges have large crossroadness values, only the first two stations (white numbers 1 and 2) fall over a manifold. This fact can be understood noting that multiple stations over the manifolds would be redundant, because they would sample trajectories intercepted by other stations. Interestingly, 173 stations were found to be necessary in order to sample the whole region. Multiplying this value for the surface of a station, i.e. $\pi \sigma^2$, we obtain that they cover only 14$\%$ of the total box surface. When considering instead a $\sigma=\delta_{OG}$, 632 stations, covering just the 3$\%$ of the total surface, were needed to sample the whole region.

 We then changed the \textit{initialization grid}, considering all the particles belonging to an ellipse of center $x_C=\pi$, $y_C=\frac{3}{2}\pi$, with semi-axes $r_x=1$, $r_y=0.5$. We note that the ellipse is thus centered exactly over a hyperbolic point. Finally, $\sigma$ is set to 0.3. The results are reported in Fig. \ref{fig:NS2Dres}C-D. Contrary to the previous case, The pattern of the crossroadness is intimately linked to the choice of the \textit{initialization grid} and also to the direction of the advection (forward or backward).

Looking at the disposition of the CR stations for the backward-in-time case it is possible to note that the first two stations are situated over the regions of highest crossroadness, and in symmetric positions in respect to the starting ellipse (dotted line). However, they do not fall exactly on the manifolds. The subsequent stations are not displaced symmetrically, and their disposition does not appear obvious. For instance, the 3rd station is on a branch of the crossroadness pattern far away from the ellipse and, interestingly, does not fall in any of the two regions with relatively high values of CR (the one at $x\simeq 5$, $y\simeq 4$ and the other at $x\simeq 1$, $y\simeq 5.5$). As in the previous case, this is due to the fact that the fluxes passing there have already been intercepted by the first two stations. The 4th station, instead, falls almost totally inside the starting ellipse. Finally, the last station (not numbered) falls in a region of higher crossroadness than the previous three.

\subsection{SVP drifters and crossroadness in the Kerguelen region}
\label{subsec:svp}

\begin{figure}[b!]
\centering
\includegraphics[width=10.0cm]{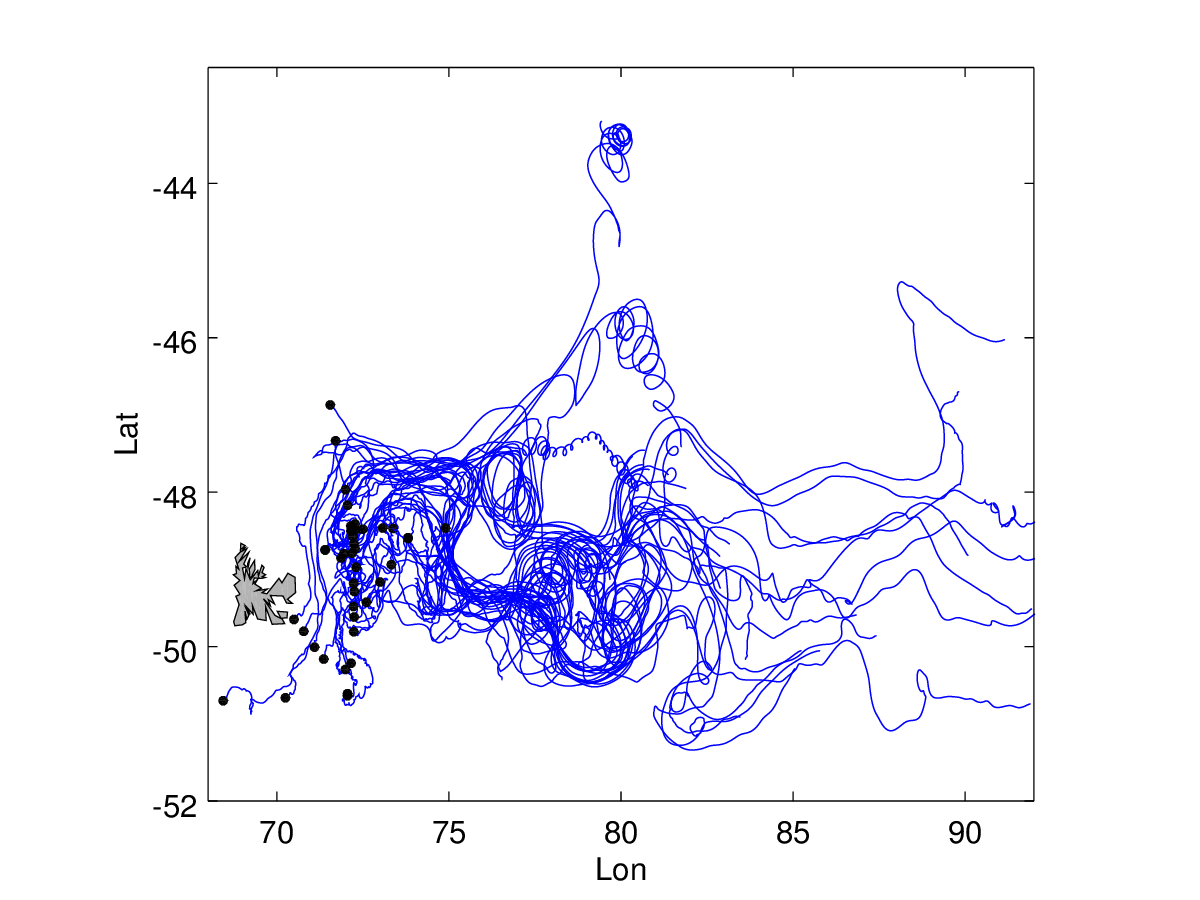}
\internallinenumbers
\caption{Trajectories of the 43 drifters of the KEOPS2 Campaign. Black dots
represent the starting positions of the drifters.}
\label{fig:drift_trj}
\end{figure}
In order to test the crossroadness and the ranking method as
defined in Subsec. \ref{subsec:CR} and \ref{subsec:CRstations} in a real oceanic environment,
we use the dataset from the KEOPS2 campaign, in which 43
drifters were released in a relatively small area (the eastern
margin of the Kerguelen plateau), approximately in the same
period of time (around the 11th of November 2011), and advected
for a similar window of time ($\uptau_D=81$ days on average) by the
currents, as shown in Fig. \ref{fig:drift_trj}. We use thus 43
real drifter trajectories.
\begin{figure}[h!]
\centering
\includegraphics[width=10.0cm]{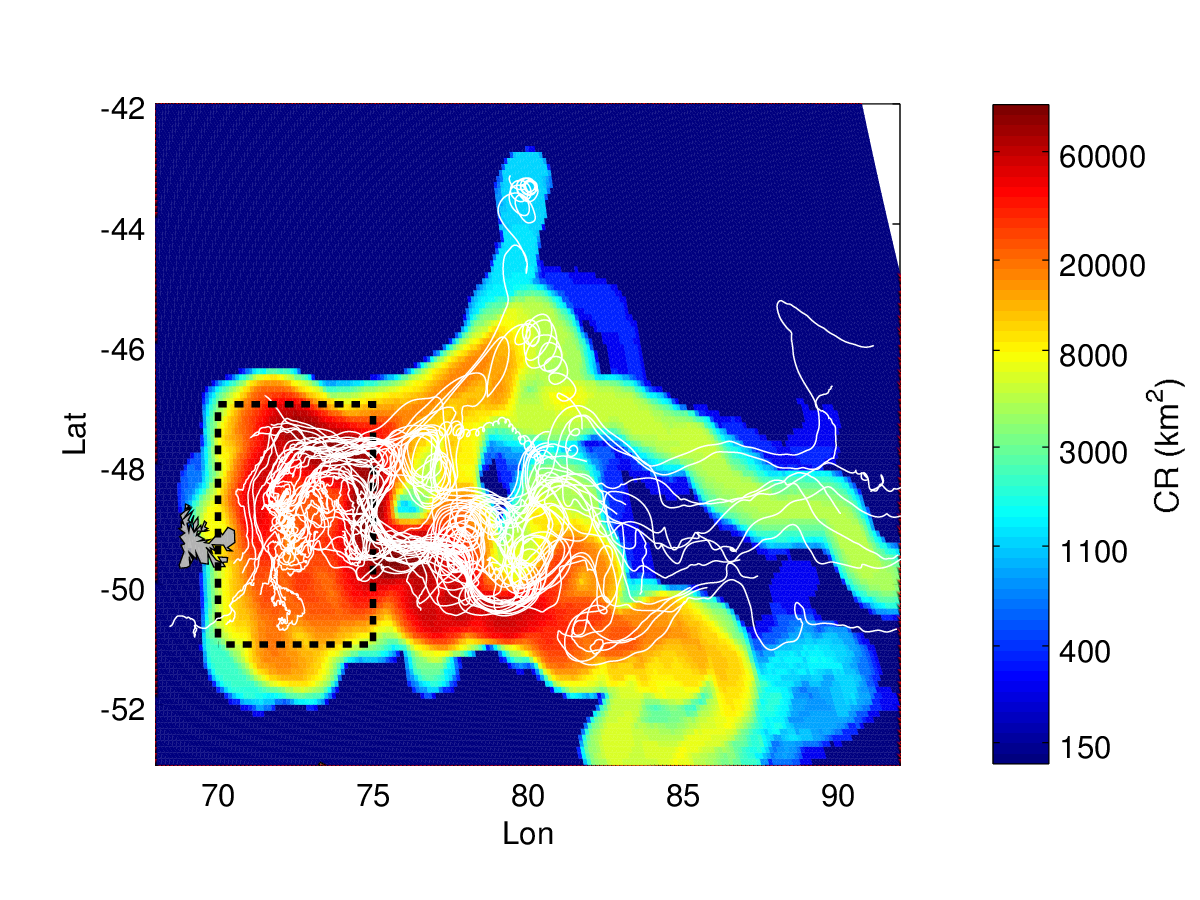}
\internallinenumbers
\caption{Crossroadness field derived from satellite altimetry ($\uptau=60$ days,
$\delta_{IG}=\delta_{OG}=0.1\degree$, $\sigma=0.4\degree$), with superimposed trajectories of SVP drifters
released during the 2011 KEOPS2 campaign. The CR field was computed advecting only the
small \emph{rectangle} (longitude: [70, 75]$\degree$; latitude: [-51, -47]$\degree$, black dotted line)
in which the drifters have been released.}
\label{fig:CRdrift_glob}
\end{figure}
\begin{figure}[h!]
\centering
\includegraphics[height=5.0cm]{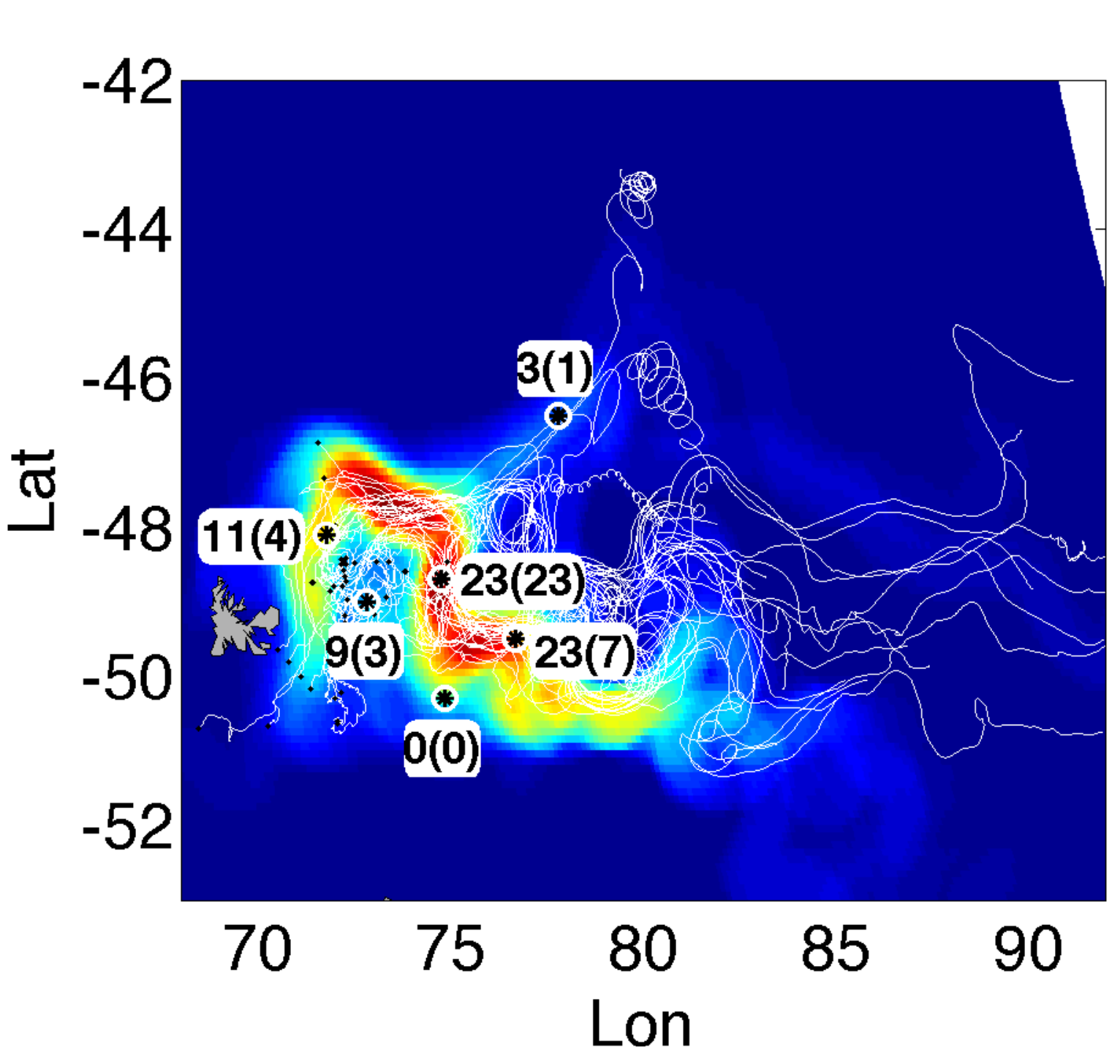}
\quad
\includegraphics[height=5.0cm]{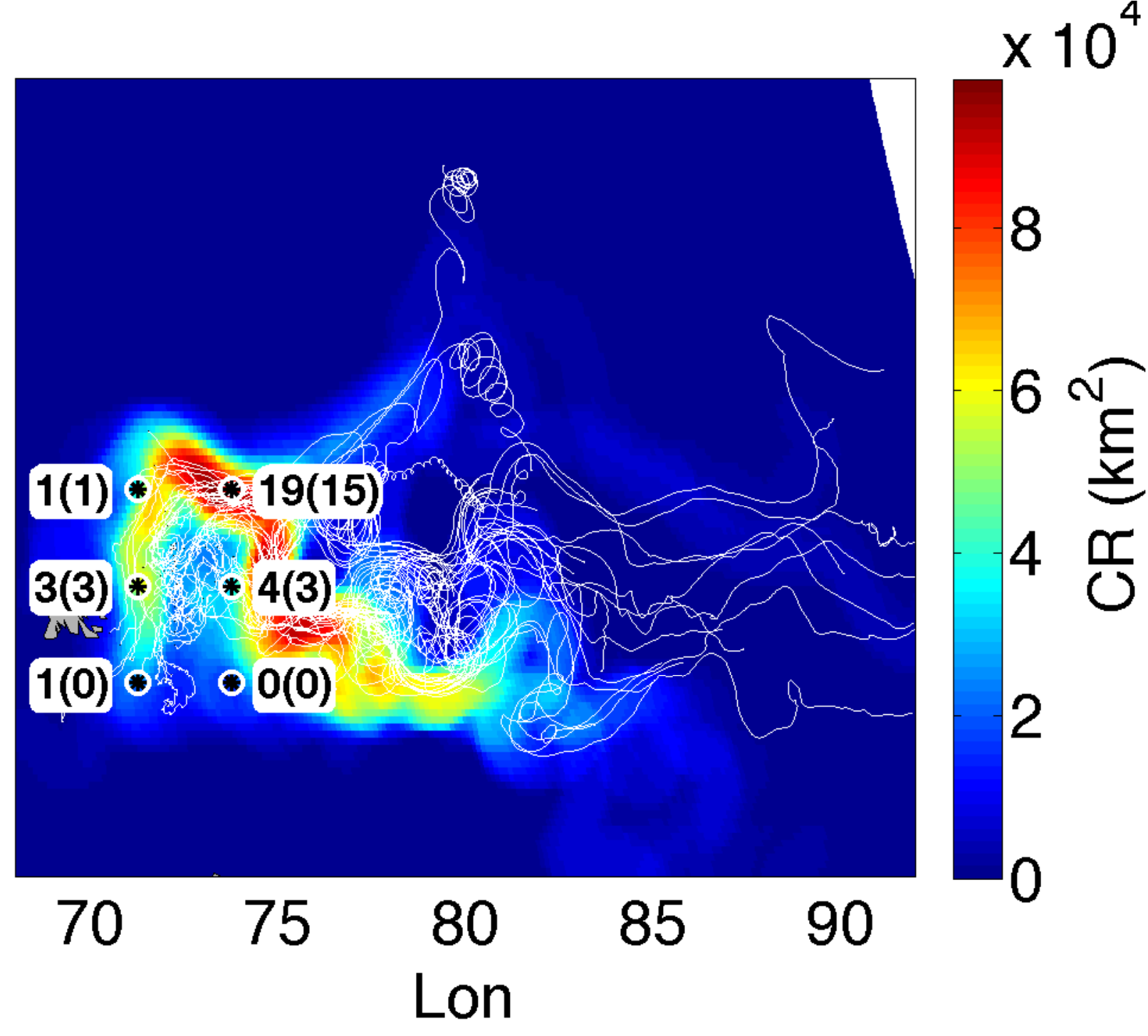}
\internallinenumbers
\caption{Left panel: first 6 CR stations (black stars) superimposed on the CR field
 computed advecting the same region as in Fig. \ref {fig:CRdrift_glob}, using the regional
 product for the velocity field ($\uptau=60$ days, $\delta_{IG}=\delta_{OG}=0.1\degree$, $\sigma=0.2\degree$).
 Right panel: same CR field, this time the black stars
identify 6 stations disposed on a regular grid. For each station, the first number identifies
the amount of total drifters intercepted, the value in brackets the number of independent
drifters. Amount of drifters intercepted by the CR stations (total and in parenthesis the
independent): 69 (38). Regular grid: 28 (22).}
\label{fig:CRstat&grid}
\end{figure}

Our aim is to compare the regions of densest passage of these real buoys with the ones predicted by the CR method, which is
based on oceanic currents estimated from satellite altimetry. Therefore, the \textit{initialization grid} consists in a series of virtual drifters displaced in the
following way: we consider the region of release of the real drifters
(between [-51, -47]$\degree$S, and [70, 75]$\degree$E). We cover it
with rows of virtual tracers, separated by $\delta_{IG}=0.1\degree$ along
the latitude. In order to preserve the same angular distance
(0.1$\degree$) among tracers of the same row, we make a latitudinal
correction on their longitudinal separation, so that
$\delta_{LON}=\delta_{IG}/\cos(LAT(row))$, as explained in Subsec \ref{subsec:CR}.
We put in each row the same number of virtual drifters. Thus, the longitude range of
the southern row (at latitude 51$\degree$S) is
[70,75.4]$\degree$. For simplicity we will denote in the
following this type of geographic region as a \emph{rectangle} of
coordinates longitude=[70, 75]$\degree$ and latitude=[-51,
-47]$\degree$. \\
We use the altimetry-derived velocities to
advect these points forward in time for a period of $\uptau=60$
days, comparable with $\uptau_D$. We compute the CR values at all points separated by the
same distance $\delta_{IG}$ on the Kerguelen region, i.e. a
\emph{rectangle} of boundaries longitude=[68, 92]$\degree$ and
latitude=[-53, -42]$\degree$, which constitutes our
\textit{observational grid}. The resulting CR field is
displayed in Fig. \ref{fig:CRdrift_glob}. We see a remarkably
good agreement between features in the drifters trajectories
and the ones in the CR field: areas of denser buoy passage
correspond with higher values of the field, suggesting a good
qualitative match between CR high values and areas of drifters
passage.
\begin{table}[h!]
\textbf{Drifters Intercepted\\\\ }
\centering
\begin{tabular}{r|c|c|cl}
&\multicolumn{2}{c}{\textbf{CR Stations}}\vline&\\
&Global product&Regional product&\textbf{Grid}\\ \hline
$\sigma$=0.1\degree&30.5(25.5)&30.0(23.5)&12.0(11.0)\\ \hline
$\sigma$=0.2\degree&53.0(34.0)&68.3(37.3)&28.0(22.0)\\ \bottomrule
\textbf{Total mean}&\textbf{41.2(30.6)}&\textbf{51.0(31.8)}&\textbf{21.6(17.6)}\\ \hline
\end{tabular}
\vspace{0.5cm} \internallinenumbers\caption{Number of drifters
intercepted (out of 43) using six CR stations (left columns) or
six stations disposed on a $3\times2$ regular grid. The value
is an average obtained changing the different parameters used
(the advection time $\uptau$, the detection range $\sigma$ and
the resolution of the \textit{initialization} and \textit{observational grids}
$\delta_{IG}$and $\delta_{OG}$, which were kept equal). For each cell, the two values correspond
respectively to the total number of drifters intercepted by the
stations and, in parenthesis, to the independent ones. First
column: global product for altimetric velocities. Second
column: regional product. Third column: stations on a regular
grid.} \label{table:1}
\end{table}
More quantitative results are obtained by considering a set of
6 best-ranked CR stations (the choice on the quantity of
stations is arbitrary) with the method explained in the
previous section. The number of real drifters intercepted by
the CR stations is compared to the result obtained by a set of
stations placed on a regular grid. This last set is constituted
by $3\times2$ stations distributed over the area of the
drifters motion, separated longitudinally by $2.5\degree$ and
$1.23\degree$ latitudinally.\\
The measure is repeated several times and for different
parameters, changing the advection time $\uptau$ (at 30, 60 and
90 days), the detection range $\sigma$ (at $0.1$, $0.2$ and
$0.4\degree$) and the resolution of the \textit{initialization} and
\textit{observational grids} $\delta_{IG}=\delta_{OG}$ (at $0.1$ and $0.2\degree$). An
illustrative example is reported in Fig. \ref{fig:CRstat&grid},
where $\sigma=0.4\degree$, $\delta_{IG}=\delta_{OG}=0.1\degree$ and $\tau=60$
days.\\
For each station, the total number of intercepted drifters is
computed, as well as the number of drifters first detected by
this station (independent ones). The results are reported in
Table \ref{table:1}. The first two columns show the number of
drifters (total and independent) intercepted with the CR
stations, the third one the values obtained with a regular
grid. The efficiency of our method is about twice the result
obtained with a regular grid. A lower value of sigma
corresponds not surprisingly to a lower number of catches for
both the CR and the regularly spaced \textit{observational grid}, but to an
improved ratio in favor of the CR network.

\subsection{Dependence of the surface monitored on the number of monitoring CR stations}
\label{subsec:sur_vs_st}

\begin{figure}[b!]
\centering
\includegraphics[width=9.0cm]{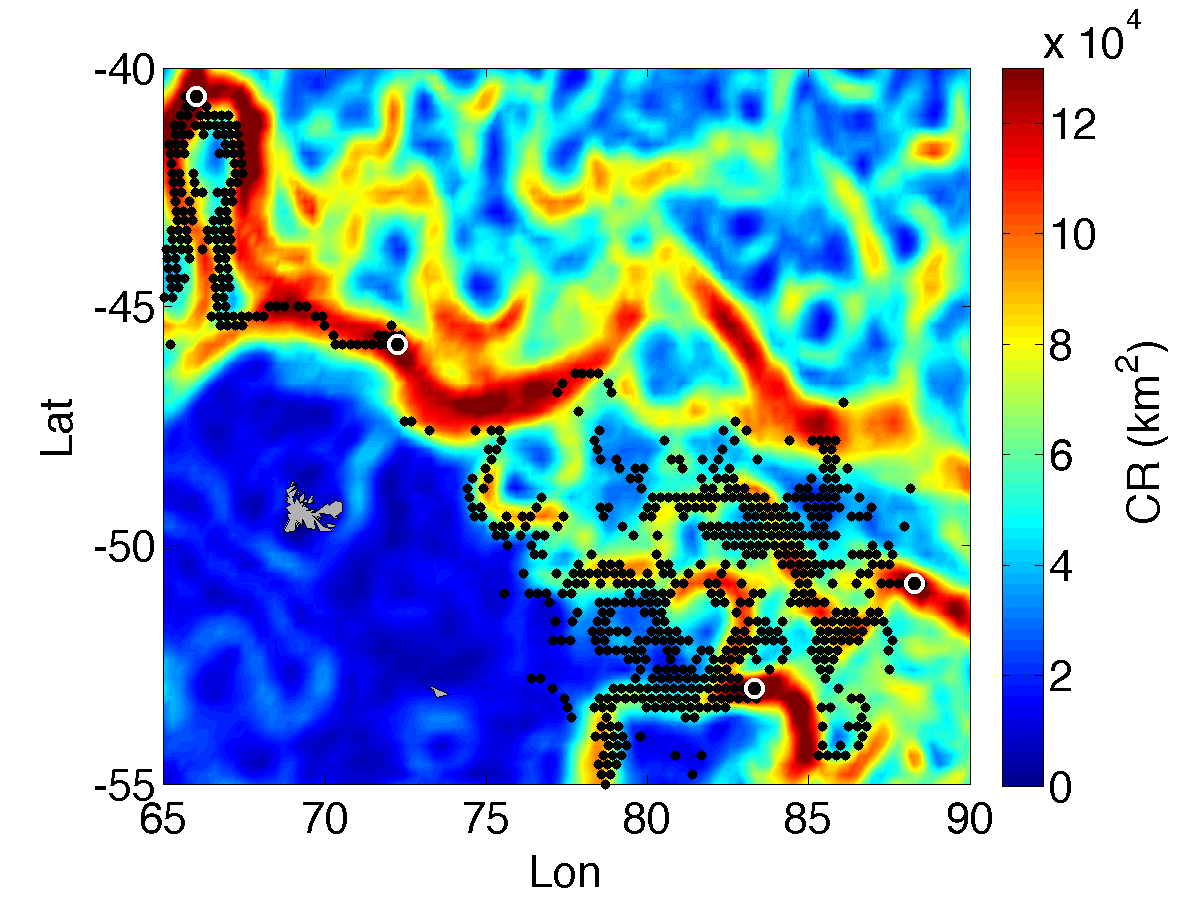}
\internallinenumbers
\caption{For this plot and the ones showed in Appendix A,
the \textit{initialisation grid} is the \emph{rectangle} between longitude=[55, 100]$\degree$, latitude=[-36, -59]$\degree$,
while the \textit{observational grid} corresponds to the domain showed,
if not specified differently. Crossroadness computed with an advection
time $\uptau=60$ days, $\delta_{IG}\,=\,\delta_{OG}\,=\,\sigma=0.2\degree$, with, superimposed,
the first four CR stations (white circles) and the surface that they control
(black dots on the corresponding \textit{initialization grid} points). Note that
some black dots can be outside the plot, since we advected
a larger region than the observational domain showed in the panel, in order to take
into account the particles upstream.}
\label{fig:CR_with_stations}
\end{figure}

\begin{figure}[t]
\centering
\includegraphics[width=7.25cm]{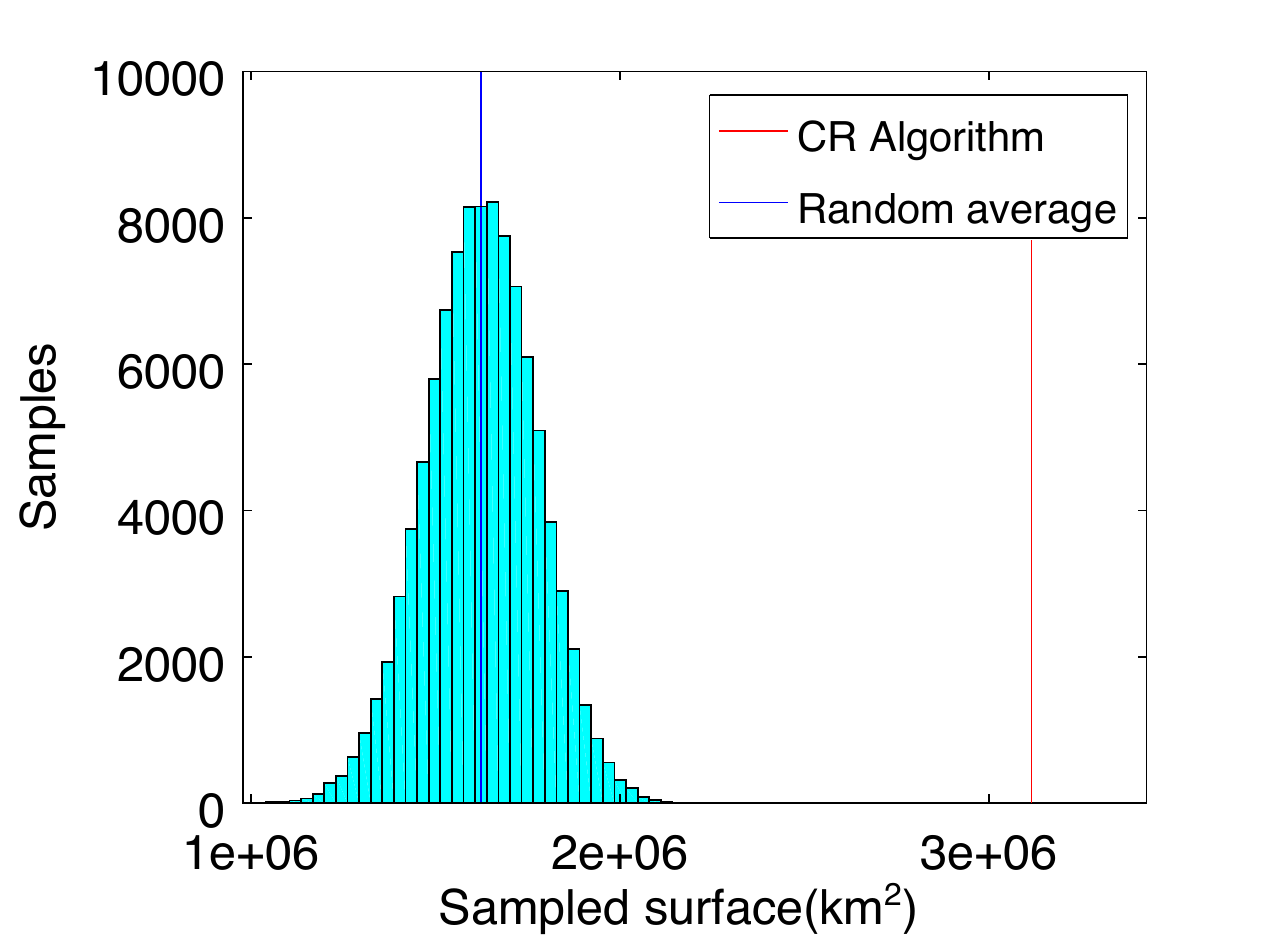}
\internallinenumbers
\caption{Histogram, computed from 100000 experiment repetitions, of the
surface detected choosing each time $N=20$ stations
randomly displaced. The vertical red line on the right is instead the surface
scanned choosing the stations with our CR algorithm. Here $\sigma=0.4\degree$. The distance between the mean value of the distribution (vertical blue line) and the vertical red line is 9.96 times the standard deviation of the distribution.
}
\label{fig:hist_20stations}
\end{figure}
\begin{figure}[h!]
\centering
\includegraphics[width=9.5cm]{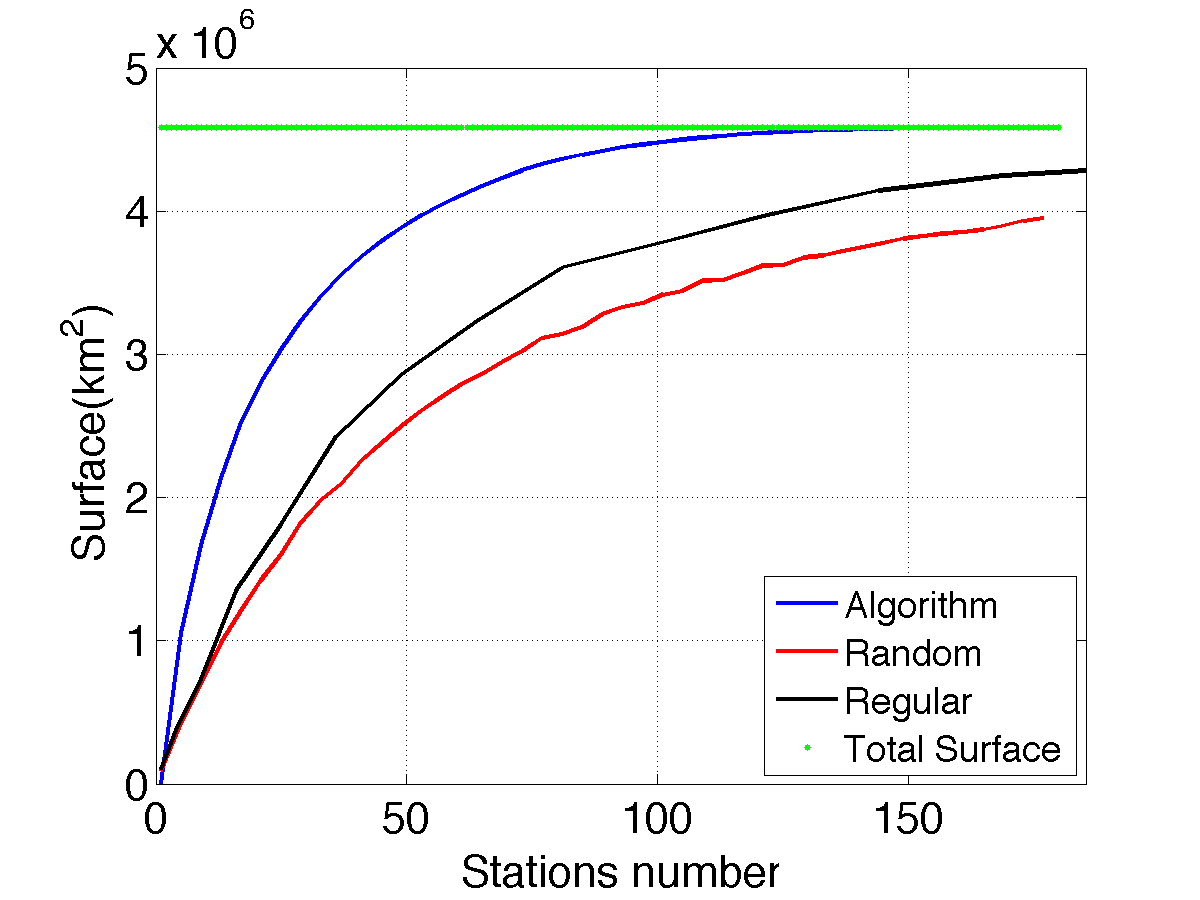}
\internallinenumbers
\caption{Surface sampled varying the number of stations considered ($\uptau=60$ days, $\delta_{IG}=\delta_{OG}=0.1\degree$, $\sigma=0.4\degree$),
chosen randomly (red line), on a regular grid (black line) and with the
CR method (blue line). Concerning the random choice and the regular grid, each value was obtained from the average of 1000 repeated measures. For the case of the regular grid, each time it was rigidly shifted along the longitude and the latitude of a random fraction of the distance between two grid points. The green line represents instead the total surface.}
\label{fig:surface(Nstations)}
\end{figure}

We present here a statistical analysis in order to assess the
efficacy of a CR based detection network
with respect to a set of stations randomly chosen or
arranged on a regular grid. The benchmark that we use is
the CR, i.e. the ocean surface crossing the CR
observing network as a function of the number of the monitoring
stations.

Figure \ref{fig:CR_with_stations} displays a map of forward
crossroadness computed around Kerguelen Island, advecting
particles starting from November 1st, 2011, for a time of 60
days. Superimposed, white circles identify the first four CR
stations. Black dots identify the monitored waters, namely the
points that will pass in proximity of one of the four stations
during the advection time $\uptau$.

First, we measure
the surface monitored by 20 randomly selected stations with a
detection range of $0.4\degree$ in a period of 60 days. The
measure is repeated $100000$ times and the results are reported
in the histogram of Fig. \ref{fig:hist_20stations}. The
vertical blue line is centered on the mean value of the
distribution, while the red one is the value of the surface
scanned with 20 stations chosen with our method. The distance
between the two measures is about ten times the value of the
standard deviation of the distribution, an extremely
significant ($p<10^{-22}$) deviation with respect to the
expected result of monitoring a larger surface with randomly
selected stations.

The measure is repeated changing the number of stations
selected, and the results are shown in Fig.
\ref{fig:surface(Nstations)}. A sampling with a regular grid is
performed as well. In this case, each measure is obtained by
rigidly shifting the grid by a random fraction of the grid step
along longitude and latitude. The use of the CR stations (blue
dots) shows a better performance compared to the regular grid
(black stars) and the random case (red circles) sampling. For instance,
to scan a surface of $3\times 10^6\:km^2$ with a detection range
of $R\sigma=40 km$, 25 stations selected with the CR method are
needed, about 55 with a regular grid, and 75 randomly chosen.

\subsection{Persistence of the monitoring network}
\label{subsec:persistence}

The calculation of the crossroadness computed in the previous
sections requires that at the moment of choosing the monitoring
stations, the velocity that will disperse the tracer in
the future is already known with good precision.
As an example, in Subsec. \ref{subsec:svp} we calculated the CR
stations used to intercept the SVP drifters using the velocity
data of the days in which the buoys were advected by
the currents. How much this impacts the CR
stations ability to intercept the maximal surface of a stirred
patch? Here we attempt to address this question by looking at the
\textquotedblleft persistence\textquotedblright of the CR
stations i.e. the ability of a CR network, using velocity of the past, to intercept a tracer dispersed in the near future.

\textit{Surface monitored using past velocity field for the
computation of the CR stations.}

In general, we define
\begin{center}
$\mathcal{T}(\mathcal{R})_{D_0\rightarrow D_f}$
\end{center}
as the whole collection of trajectories $\mathcal{T}$ generated
from the advection of all the points regularly
initialized over a region $\mathcal{R}$, from the day
$D_0$ until $D_f$. In general, from a set of trajectories
$\mathcal{T}(\mathcal{R})_{D_0\rightarrow D_0+\uptau}$, we can
compute an ensemble of CR stations
$\mathcal{S}(\mathcal{R})_{D_0\rightarrow D_0+\uptau}$, as
explained in Subsec. \ref{subsec:CRstations}, that will be for
construction the best choice in order to scan $\mathcal{T}$.

When the velocity field between the day $D_0$ and $D_0+\uptau$
is not known, we can use the stations computed with the
velocity field in a time interval previous to $D_0$, e.g.
$\mathcal{S}(\mathcal{R})_{D_0-\uptau \rightarrow D_0}$ to
monitor the area, under the assumption that if the flow does
not change much in an interval of time of the order of $\uptau$
the optimal stations will maintain approximately the same
positions.

\begin{figure}[h!]
\centering
\includegraphics[width=5.5cm]{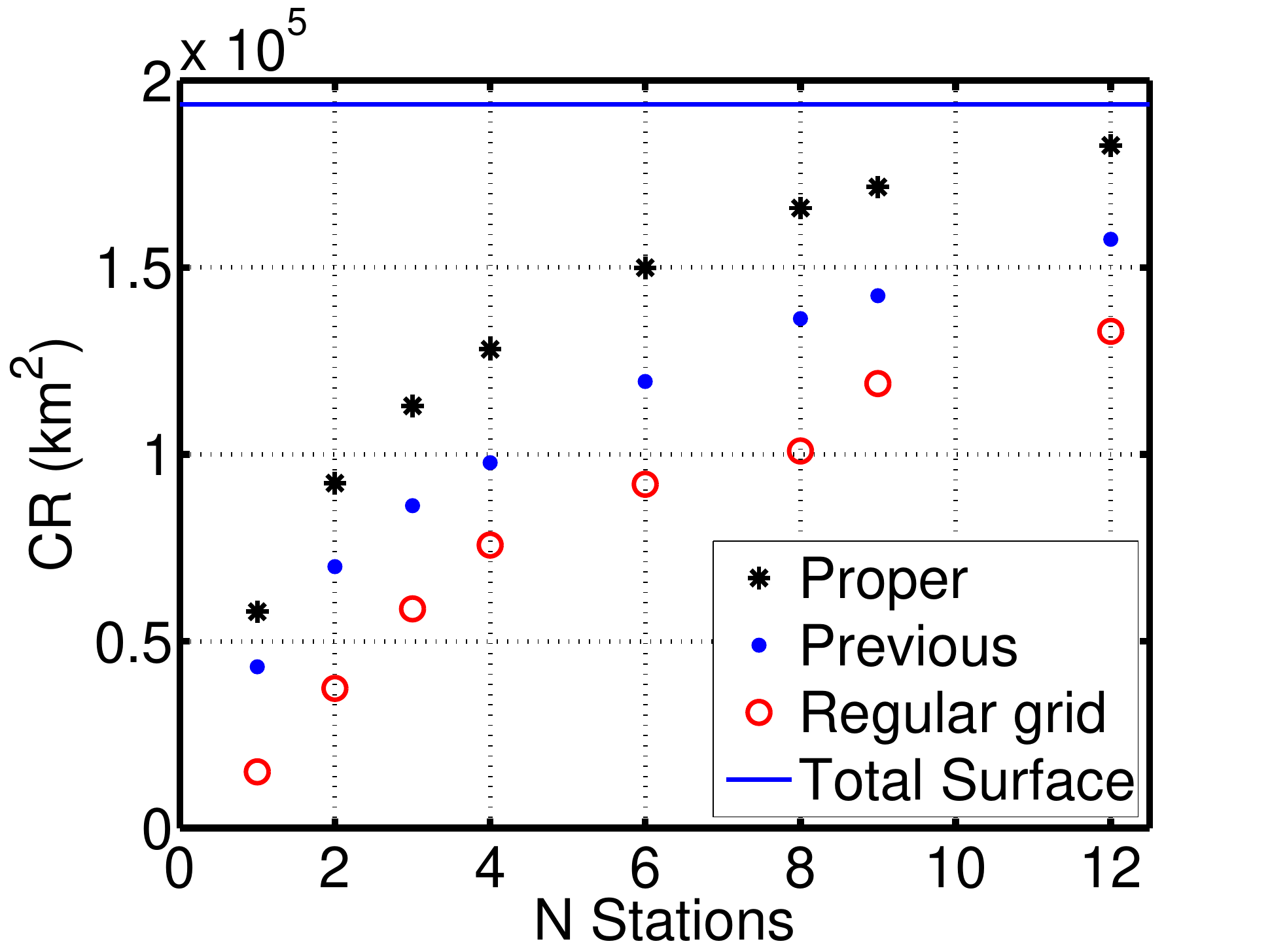}
\quad
\includegraphics[width=5.5cm]{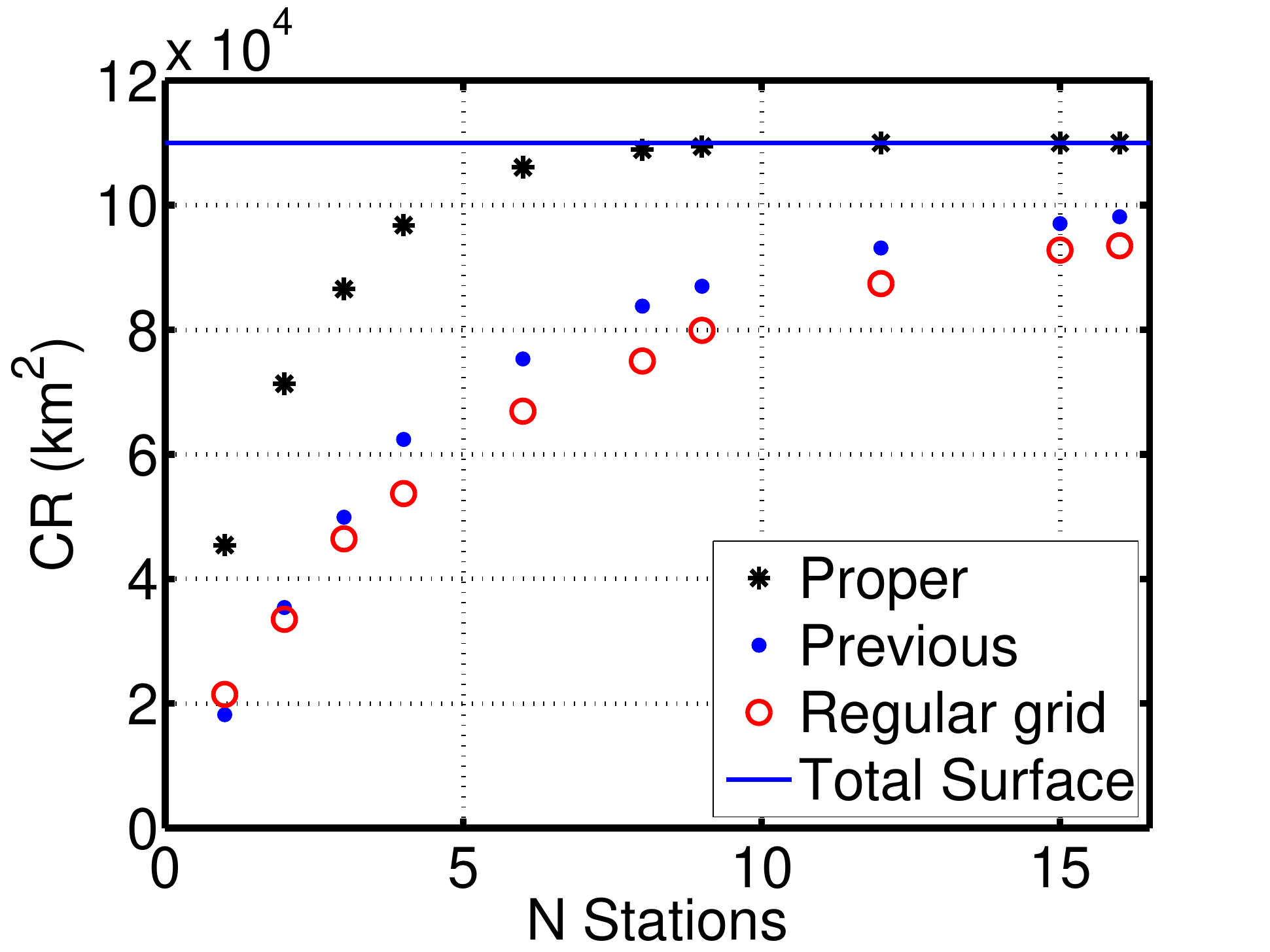}
\internallinenumbers
\caption{CR, i.e. surface monitored as a function of the number of stations. Each point is the average obtained changing $D_0$ over the first day of each month of 2011.
Black stars represent the value obtained using the stations computed
with the proper velocity field, i.e. $\mathcal{S}(\mathcal{R})_{D_0\rightarrow D_0+\uptau}$,
while blue dots the ones computed with the previous $\uptau$ days, (i.e.
$\mathcal{S}(\mathcal{R})_{D_0-\uptau\rightarrow D_0}$).
Red circles are the values obtained with a regular disposition of the sampling sites.
Left panel: $\mathcal{R}\,=\,\mathcal{R}_P$ (plateau region). Right panel:
$\mathcal{R}\,=\,\mathcal{R}_T$ (turbulent area). $\delta_{IG}\,=\,\delta_{OG}\,=\,0.2\degree$,
$\uptau=60\,$days, $\sigma\,=\,0.2\degree$.
}
\label{fig:an42}
\end{figure}
Thus, in this section we use the velocities between
$D_0-\uptau$ and $D_0$ to compute a set of stations
$\mathcal{S}(\mathcal{R})_{D_0-\uptau\rightarrow D_0}$. We use
then the latter to monitor
$\mathcal{T}(\mathcal{R})_{D_0\rightarrow D_0+\uptau}$ and see
how many trajectories (i.e. sea surface)
they intercept.

We take as $D_0$ the 1st January, 2011 and $\uptau=$ 60 days.
We change the number of stations considered between 1 and 16,
analogously to what has been done in Subsec.
\ref{subsec:sur_vs_st}. For comparative purposes, we
also measure the surface intercepted with
$\mathcal{S}(\mathcal{R})_{D_0\rightarrow D_0+\uptau}$, and
then with a regular disposition of the stations. We then repeat
the procedure using as $D_0$ the 1st February 2011, then the
1st March 2011 and so on, until the 1st of December 2011 and we consider the average of the 12 values obtained. In
this way we obtain a more consistent statistics.

The results are reported in Fig. \ref{fig:an42}. We consider
two different monitoring areas: the first one is the drifters
release area, defined in Subsec. \ref{subsec:svp}, situated
mainly on the Kerguelen plateau and in which the bathymetry
seems to affect the circulation pattern, making it more
persistent in time. We will refer to it as $\mathcal{R}_P$. We
then consider a more turbulent region situated offshore from
Kerguelen, in which the current field should not be as
constrained by the shallow shelf structures as in the former
case, and in which the mesoscale structures affect deeply the
variability of the currents (\citealt{park2014polar}). We take
this turbulent region $\mathcal{R}_T$ to be the
\emph{rectangle} with longitude between $80$ and $83\degree E$,
and latitude between $47$ and $50 \degree S$.

Concerning the plateau region (left panel in Fig.
\ref{fig:an42}), we see a stronger performance of the CR
stations compared to the regular grid, even if they are
computed with the trajectories of two months before. E.g., in
order to monitor a surface of $100000\,km^2$, we need 8
stations disposed regularly, while only 4 if we consider the
stations obtained from the advection of the previous 60 days,
and 3 if we consider the advection from time $D_0$ (i.e.
the velocity field simultaneous to the surface
advection).

The results are worse but surprisingly consistent for
the turbulent region scenario (right panel in Fig. \ref{fig:an42}), with about $25\%$
less CR stations needed than in the regular case. The
analysis in this section is completed in Appendix A, in which
the detection power of the CR stations is assessed against
further changes in the dates used for the velocity field.

\subsection{Identification of a source region}
\label{subsec:source_region}

In the previous section we used CR for
intercepting a tracer stirred from a given region. Here instead
we study another typical problem arising when studying
dispersion, namely the identification of the most important
source regions connected by the circulation to a given target
area.

This occurs, for instance, when we aim to determine the key
regions that can affect vulnerable marine protected
areas downstream, or in the identification of nutrient sources
feeding a biogeochemical active region (\citealt{ciappa2014oil,suneel2016backtrack}).

In order to showcase this application, we considered the area
studied in the previous section, i.e. the Kerguelen plateau,
that recent studies have stressed as a natural source of iron
supply that sustains the primary production in the zone
situated to the east of the island
(\citealt{d2015biogeochemical, blain2008distribution,
christaki2008microbial}).

Large areas of the Earth oceans present waters with high
quantities of nutrients, but low concentration of chlorophyll
(HNLC). This is generally due to the absence of some
micronutrients that act as limiting factors. In many cases one
of the main constrains to the presence of chlorophyll is the
low concentration of bioavailable iron. In this regime,
injection of this micronutrient fuels the primary
production
(\citealt{boyd2007mesoscale,lam2008continental,martin2013iron}).
In recent years, different studies have underlined the
importance of continental margins as a subsurface source of
iron that can thus fed the phytoplanktonic bloom in the waters
downstream (\citealt{lam2008continental}).
\begin{figure}[t!]
\centering
\includegraphics[width=5.5cm]{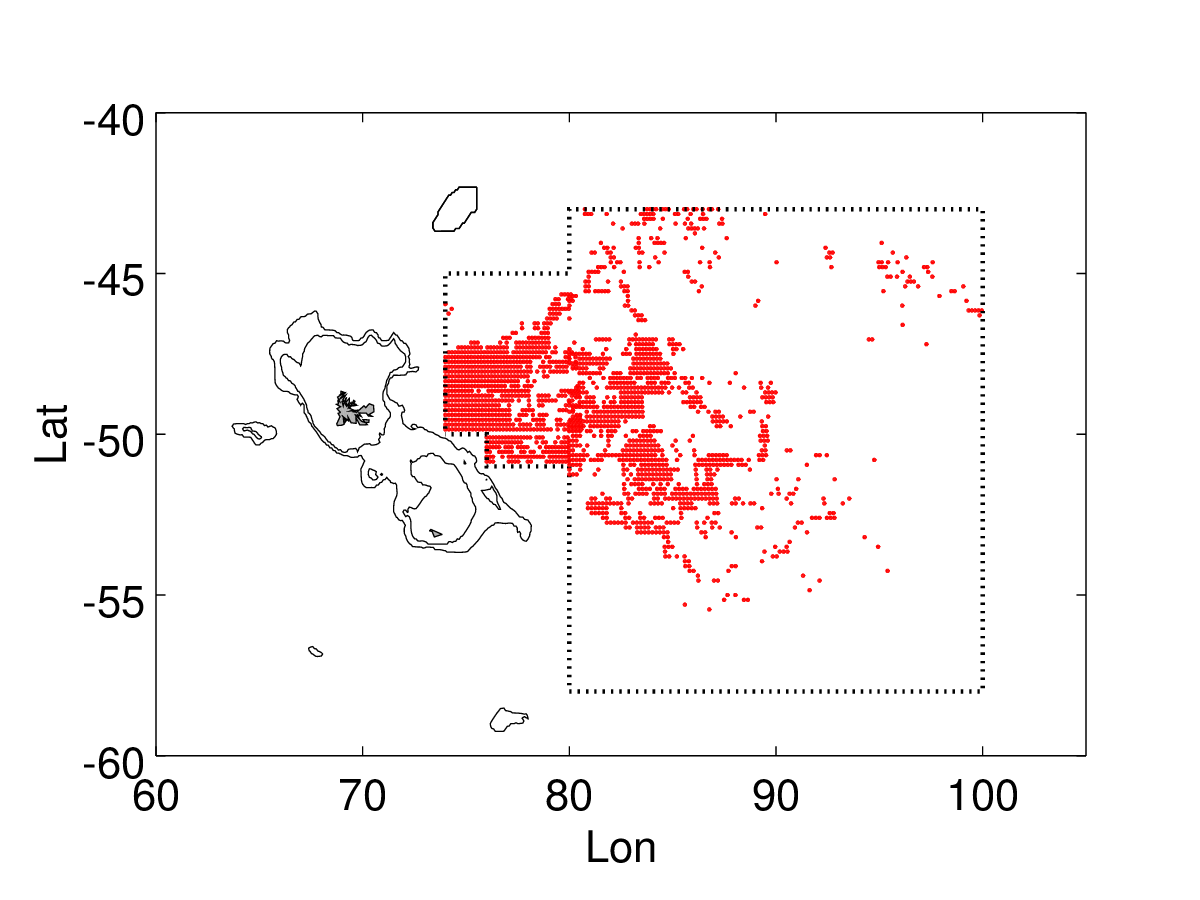}
\quad
\includegraphics[width=5.5cm]{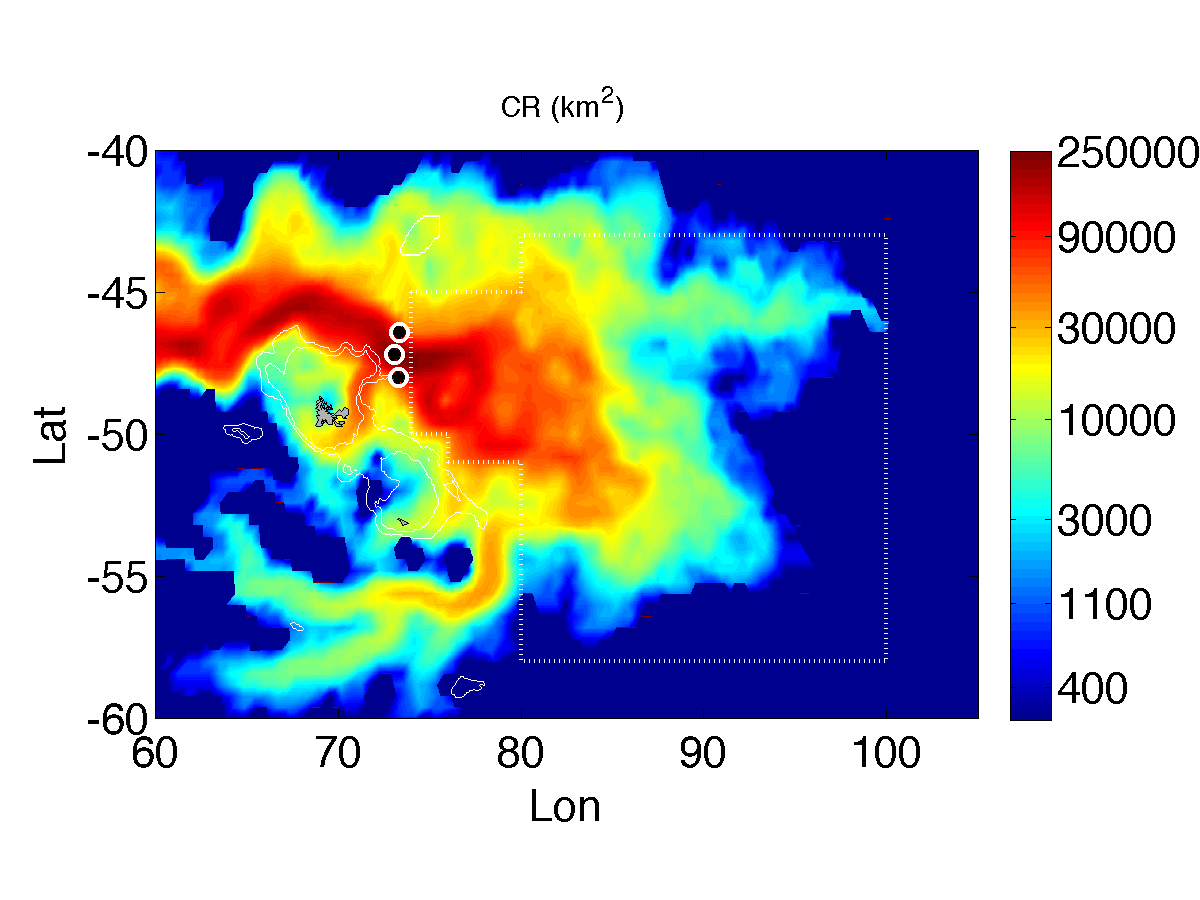}
\internallinenumbers
\caption{Left panel: Kerguelen plateau with the bloom area delimited by black
dotted lines. Bathymetric lines at 500 and 1000 meters. The set of initializing points (shown in red) are
selected by choosing the pixels that supported at least one value of chlorophyll
greater than $1\,mg/m^3$ between November and December 2011 and are showed in red.
Each initialization point is assumed to carry an amount of water given by a surface
$\Delta=R^2\delta_{IG}^2$ (when $R$ is the Earth radius and $\delta$ is in radians) for
$\delta_{IG}\,=\,0.15\degree$.
Right panel: CR computed backward (in log scale), with advection time
$\uptau\,=\,180$ days, from the red points showed on the left panel,
that cover a total area of about $517000$ km$^2$. Black dots represent the
first three source stations computed with the CR algorithm described above.
Detection range $\sigma\,=\,0.4\degree$}
\label{fig:Kerg_bw_area}
\end{figure}
Among them, the Kerguelen plateau is one of the clearest region
to show this biogeochemical dynamics and its importance for
primary production is now well established. Nevertheless, the
hotspots of the continental shelf that may act as main iron
sources are still unknown. Here we use the crossroadness
computed backward in order to address this question.

Since we want to determine the possible iron sources on the Kerguelen plateau, delimited by the bathymetric line of 1000m, we examine only the region situated eastward to this contour. This is approximated by considering only a region delimited by the perimeter reported in Fig. \ref{fig:Kerg_bw_area} (left panel, black dotted line). Furthermore, in
order to consider only the waters that support a springtime
bloom, we analyze satellite-derived images of the chlorophyll
patch of spring 2011. We therefore consider as starting points
only the pixels that, belonging to the area enclosed by the perimeter, present at least one value of
chlorophyll concentration larger than $1\,mg/m^3$ during the
peak of the bloom of November and December 2011. Note that, in this way, we do not have an \textit{initialization grid}, but a set of initialization points. The selection is reported on the left panel of Fig. \ref{fig:Kerg_bw_area} (red points). These
points are advected backward for a period of time $\uptau=180$
days, and the crossroadness field is reported in Fig.
\ref{fig:Kerg_bw_area} (right panel). This map highlights a
strong passage of water coming from the northern part of the
plateau, underlining the central role of the Antarctic
Circumpolar Current (ACC) in the advection of waters in
proximity of the island platform, and shows also the Southern
ACC Front (SACCF) that passes into the Fawn trough. We then use
the ranking algorithm in order to locate the most important
passage points which feed water to the blooming region. The
first three CR stations are all in proximity of the northern
part of the plateau, meaning that in the 180 days before the
bloom, the largest part (about 70 $\%$) of the water that will sustain primary
production, passed through these three points.
This analysis locates the Gallieni Spur as a strong candidate where
to search for iron sources of the Kerguelen bloom. We
note that in general our CR algorithm does not really determine
\emph{source} regions, but rather regions of dense trajectory
passage. But since half lifetime of iron in these waters is of
the order of about two-three weeks (\citealt{d2015biogeochemical}), for the value of $\uptau$ used our method in this
case should also locate source regions.

\section{Discussion and perspectives}
\label{sec:discussion}

Studying dispersion problems, in particular at scales
 of $\thicksim
10-100\,$kms, (\citealt{boyd2007mesoscale,olascoaga2013drifter,mahadevan2016impact}), is a central issue in many oceanographic problems (\citealt{bellingham1996optimizing,mooers2005cross,bradbury2008global,rossi2014hydrodynamic,dubois2016linking}), whose aim is to achieve good characterization, managing and protective strategies of marine areas and resources. We stress their importance, in particular, for Marine Spatial Planning (MSP). The goal of MSP is indeed to identify the peculiarities, from an ecological or societal point of view (\citealt{crowder2008essential}), of the different marine regions, and to map their spatial and temporal distributions in order to manage them in a sustainable way (\citealt{ansong2017approach,ehler2018marine}). MSP has become, in the last 20 years, a fundamental process in sea management (\citealt{douvere2008importance,munoz2015implication}) and is expected to play an increasing important role in the future (e.g., \citealt{qiu2013emerging,magris2014integrating}). An essential component of MSP is the necessity for an effective modeling and monitoring strategy (\citealt{ehler2017guide,ehler2018marine}), a question intimately related to the dispersion of tracers.

In this context, we can identify two main questions.\\
From one side, we have the case of a passive tracer, advected
by the currents. The chaotic and turbulent dynamics
characterizing the ocean circulation can disperse the patch and
make it spread over a large area (say, of size larger
than $100$ km) within a short time (days to weeks). In this
case, a recurrent problem is to locate the sites where to
deploy observing stations, capable of monitoring or collecting
the dispersed tracer (\citealt{addison2018new}).\\
A second class of problems concerns the case of a sensible
region that is influenced by the circulation upstream (\citealt{viikmae2011spatial,delpecheellmann2013investigating,delpecheellmann2013using,soomere2015towards}). In this
case, the identification of the \textquotedblleft
sources \textquotedblright that may affect and spread all over
the target area is important for vulnerability assessment (\citealt{halpern2007evaluating}).\\
Current methods, mainly based on Lagrangian advection of
particles (\citealt{dovidio_mixing_2004,mancho_tutorial_2006}), concern mostly the identification of coherent
regions with minimal transfer of water toward the environs (\citealt{haller2001lagrangian,mancho2004computation,shadden2005definition,beronvera2008oceanic}).
Nevertheless, the knowledge of the spots with enhanced exchanges amongst a flow system is a central issue in
dispersion problems (\citealt{ser2015most,monroy2017sensitivity}). The latter question has been recently
addressed in the study of Lagrangian Flow Networks (\citealt{ser2015flow,lindner2017spatio,rodriguez2017clustering,fujiwara2017perturbation}), in
particular with the concept of Most-Probable-Path\textendash betweenness (\citealt{ser2015most}). This diagnostic
identifies the choke points in the topology of a flow
system. These, nonetheless, are not necessarily spots of major
water passage.\\
Furthermore, to our knowledge,  current notions do not solve the
problem of displaying and sorting a series of stations
in order to answer the two questions mentioned above.\\
To address these issues, we introduced here a new
diagnostic, the crossroadness, which measures the water surface
flowing in the neighborhood of a point in
a given time window. This has permitted us to develop a ranking
method that estimates the places where the majority of the flux
passes, and that at the same time sees waters coming from
different locations.\\
This allowed us to design an optimal monitoring system, because
each station identified in this way intercepts independent
patches of water. We stress that this independence is important for retrieval strategies,
for example in the recovering of a contaminant. In fact, in that case the
series of the recuperation stations has to be set so that each
of them recaptures a different portion of the pollutant that is
dispersed. Thus, there is no interest that a station
monitors again some waters that have already been intercepted
upstream by another one. The same logic is valid in sampling
strategies, in which the analysis of the largest surface
possible is preferred, and in which sampling twice the same
portion of water may be a waste of resources.\\
Reversing the analysis backward in time we can instead
quantify, for each point of the domain, the amount of
surface that, passing nearby, will feed a target region. In
this case the ranking method identifies the major
\textquotedblleft source\textquotedblright points from which
the water distributes over a vast surface, with each source
\textquotedblleft irrigating\textquotedblright different areas.
The independence of the destinations allows us to maximize the
surface covered with a minimal identification of source
stations. This is an important factor for the assessment of
vulnerable points whose contamination can lead serious damages:
for instance, for the protection of Marine Protected Areas (\citealt{rengstorf2013high,ciappa2014oil}) or hotspots of
biological importance (\citealt{hobday2014identification}), like a region with a
recurrent bloom that sustains the local trophic chain (\citealt{lehahn_stirring_2007,mongin2008seasonality,d2015biogeochemical}).
In those cases, it is central to determine the main sites
upstream feeding the whole areas.\\
We first explored the properties of the crossroadness using a steady velocity field obtained from the Navier-Stokes equation (\citealt{boffetta2002intermittency,hairer2006ergodicity}), characterized by eddies surrounded by hyperbolic points and manifolds. We showed that the crossroadness is strongly linked to the choice of the \textit{initialization grid} region. Furthermore, the disposition of the CR stations is not obvious, since they do not necessarily fall on Lagrangian Coherent Structures identified by ridges of Lyapunov exponents, neither their disposition is symmetric, even if the flow stream analyzed is stationary. Interestingly, we found that the stations necessary to monitor all the box cover just the 3\% of its surface.\\
We then applied the previous concepts to the Kerguelen region by considering satellite derived velocity fields. We validated the forward-in-time case by analyzing the
trajectories of 43 SVP drifters from KEOPS2 campaign. We showed
that 6 stations computed with the crossroadness would have been
able to intercept on average about the double of the drifters
captured with a regular grid, using the same number of
detecting sites. Interestingly, the ratio seemed to improve when
diminishing the detection range. We then studied the
persistence of an optimal crossroadness network, by looking at
how its intercepting capacity degrades when the network is
computed from a velocity field previous to the one that
disperses the tracer. Even in this case, the crossroadness
stations show better performances than regular grids. This is
valid also when we consider a region with stronger turbulence.
These facts demonstrate how stations computed from past
velocity fields can be applied to future circulation patterns,
validate furthermore the algorithm proposed and show its robustness when applied to an oceanic environment.
It is in fact presumable that, for growing levels of turbulence, the performances of the CR stations, those of a regular grid or a random disposition of the stations would tend to coincide, due to the chaotic activity. However, even when considering a zone with a very strong turbulence (the Kerguelen region), the algorithm showed better performances than a regular grid, meaning that the levels of turbulence activity in ocean do not affect the capacity of the algorithm. The effectiveness of the algorithm does not rely either on a strong detecting power of the stations employed, but on the contrary it improves when diminishing it. The only limitation is the fulfillment of the condition stated in Subsec \ref{subsec:CR}, i.e. $\sigma\geq\delta_{IG}$, which could lead potentially to a decrease in the algorithm performance if the detection range considered is smaller than the resolution of the velocity field.
We note that our methodology determines
inflow and outflow pathways through fixed observation stations.
This choice fixes a preferred reference frame. This is at
variance with methodologies addressing a different set of
problems for which frame-invariance is usually a requirement,
such as the computation of Lagrangian coherent structures
(\citealt{Hadjighasem2017critical}).
We use the backward-in-time method to analyse the Kerguelen
spring primary production during November-December 2011,
showing that about the 70\% of the waters that supported the
bloom had passed in the vicinity of just 3 sites on the
Kerguelen plateau during the previous 6 months, in proximity of
the Gallieni Spur.\\
In our analyses we focused on the properties of the crossroadness considering two-dimensional dynamics. 2D turbulence is in fact present in nature over a large range of scales in which the ratio of lateral and vertical length is very large (\citealt{kraichnan1980two,tabeling2002two,boffetta2012two}). When applying these concepts to the oceanic cases, we considered periods of advection sufficiently small in order to neglect vertical velocities. Furthermore, several relevant tracers (in a first approximation) such as plastic, oil and chlorophyl are present almost only in the upper ocean layers, and for their study the 2D approximation can be considered very robust.\\
We note also that all the analyses provided here can be integrated in a three-dimensional environment, delineating interesting perspectives for future studies.

We note that in a recent paper, \citet{rypina2017trajectory}
introduced a \textquotedblleft mixing potential\textquotedblright approach which exploits ideas
similar to our crossroadness. The main difference is
that the \textit{mixing potential} is a Lagrangian quantity
attached to each fluid parcel, whereas our
crossroadness uses Lagrangian trajectories but is
assigned to each fixed location in space. Thus, whereas the diagnostic
in \citet{rypina2017trajectory} may be more appropriate to
assess mixing, our approach aims to
deploy observation networks and identify sources of transported
substances.

There are then several other cases looking interesting for future applications of the crossroadness and its
ranking method, and we list here some.
For the dispersion of pollutants, the forward in time crossroadness
allows us to estimate the most important points in which position
a fix station in order to retrieve the contaminant. The
aforementioned method can also be used for sampling strategies
in order to maximize the surface intercepted and the
probability of encountering elements of interest and thus to improve the cost-effectiveness quality (\citealt{elliott1996need,elliott2011marine}). In search and
rescue operations, if the exact missing point is lacking, and
the information available is just on the area of disappearance,
computing the forward CR could establish optimal
observing stations to look for the lost target. Regarding the backward calculation, this can be
used for prioritizing survey locations upstream to vulnerable
regions (like Marine Protected Areas), or identifying most likely hotspots
close to the shore from which fish larvae may span to a large
recruiting area. Our diagnostic provides a direct and simple way to sort a series of stations in order to survey with good performances even very turbulent regions, taking furthermore explicitly into account the detection range.

\section*{Acknowledgements}
This work is a contribution to the CNES/TOSCA project LAECOS
and BIOSWOT, and it was partly funded by the Copernicus Marine
Environment Monitoring Service (CMEMS) Sea Level Thematic
Assembly Centre (SL-TAC). C.L and E. H-G. acknowledge support
from Ministerio de Econom\'{\i}a y Competitividad and Fondo
Europeo de Desarrollo Regional through the LAOP project
(CTM2015-66407-P, MINECO/FEDER). The authors thank also
Isabelle Pujol and Malcolm O'Toole for their helpful advices.
We thank furthermore Guido Boffetta for his help with the 2D Navier Stokes model.

\section*{Appendix A. Temporal persistence of the CR stations along the year}
\label{sec:AppA}

\renewcommand{\theequation}{A.\arabic{equation}}
\setcounter{equation}{0}  
\renewcommand{\thefigure}{A.\arabic{figure}}
\setcounter{figure}{0}  
\renewcommand{\thetable}{A.\arabic{table}}
\setcounter{table}{0}  

In this Appendix we extend the analysis of Subsec.
\ref{subsec:persistence} on the possibility of using known
velocity fields from the past to obtain sets of CR stations
able to monitor transport by future velocity fields.\vspace{0.25cm}

\textit{Drifters intercepted with stations computed along the
year.}

\begin{figure}[t!]
\centering
\includegraphics[width=5.5cm]{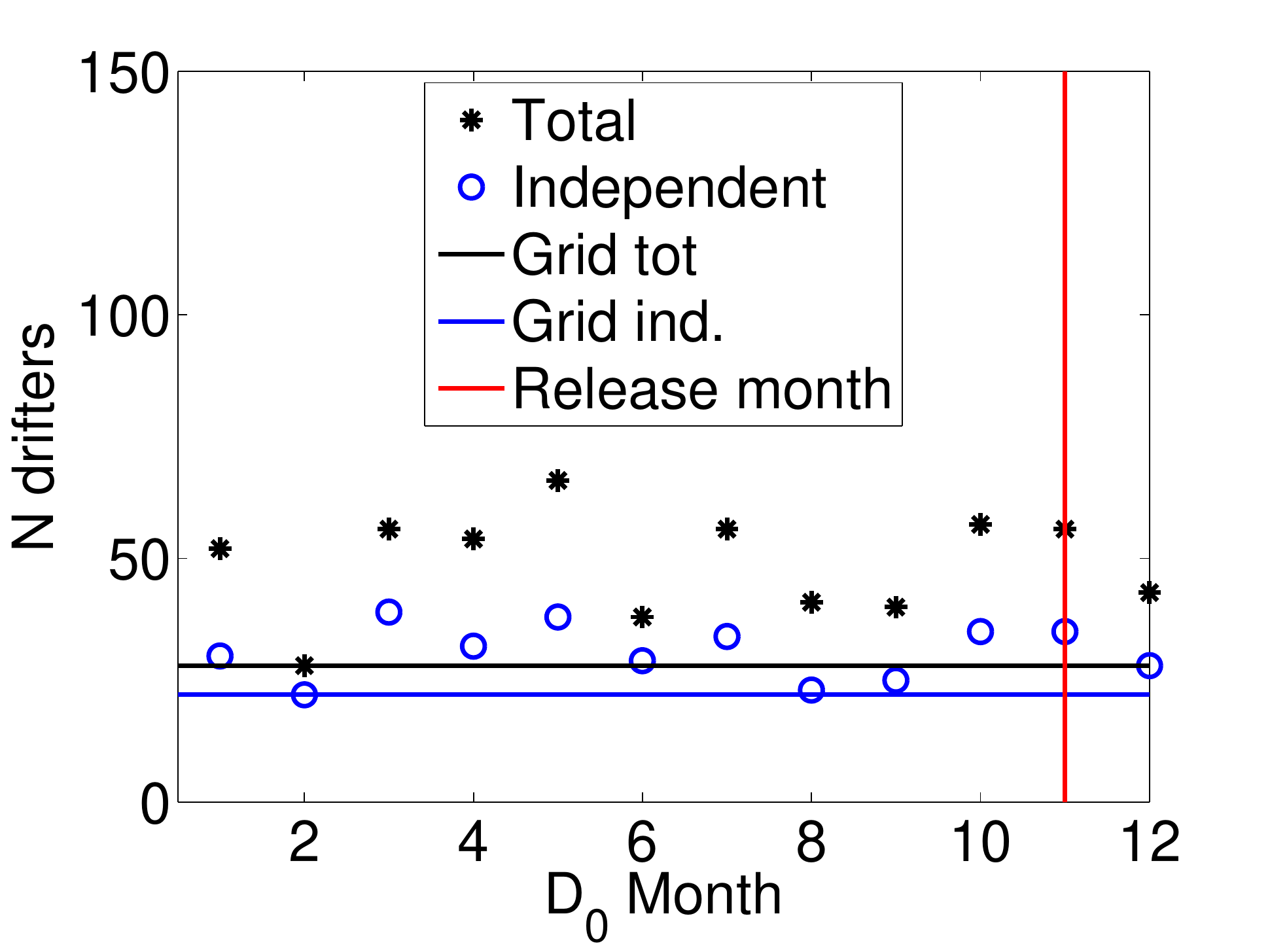}
\quad
\includegraphics[width=5.5cm]{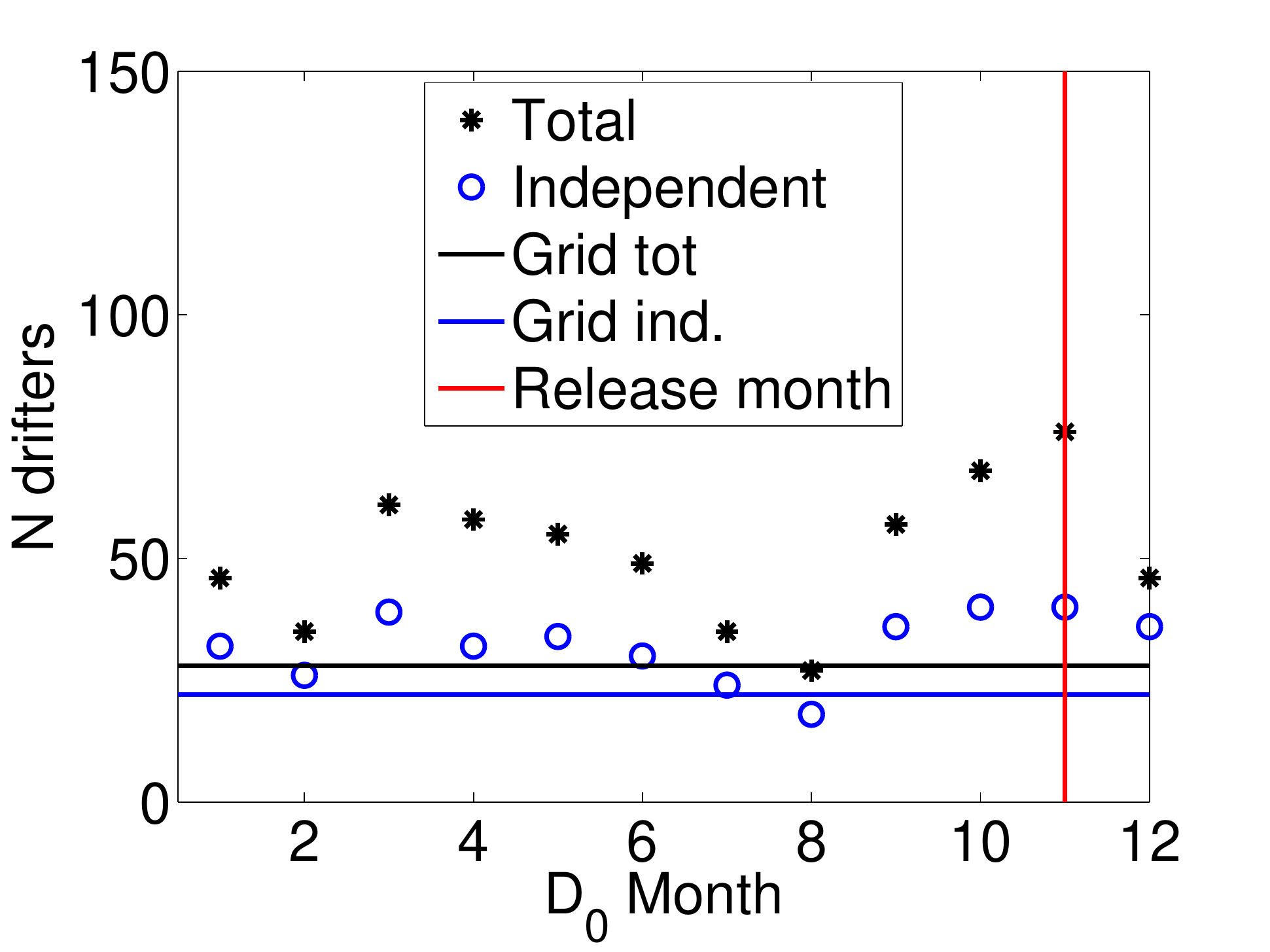}
\internallinenumbers
\caption{Number of drifters intercepted using the different
months along the year. $\delta_{IG}\,=\,\delta_{OG}\,=\,\sigma=0.2\degree$.
Left panel: $\uptau=30\,days$. Right panel: $\uptau=60\,days$
Horizontal lines: values obtained with a regular grid, total
(black) and independent (blue line).
Note that the release period of the drifters is November 2011.}
\label{fig:an33}
\end{figure}

The SVP drifters release region $\mathcal{R}_P$ is advected
taking as starting day $D_0$ January, 11th, 2011, for a period
$\uptau$ of 30 and 60 days, from which 6 CR stations were
calculated. They are then used to see how many real drifters
from the dataset released on November 2011 they would have
intercepted. The computation is repeated changing $D_0$, using
each time a different month until December, 2011. The results,
reported in Fig. \ref{fig:an33}, show a better performance of
the CR stations compared to the regular grid, along all the
year, with a number of drifters intercepted always higher
except for one case (August 2011, $\uptau\,=\,60$ days).
The case of 60 days advection presents a linear decrease
of drifters intercepted for calculations using the three months
previous to November, and then a regular increase again,
showing a sort of annual cycle, while the 30 days results  shows a more irregular trend.

\begin{figure}[t!]
\centering
\includegraphics[width=5.5cm]{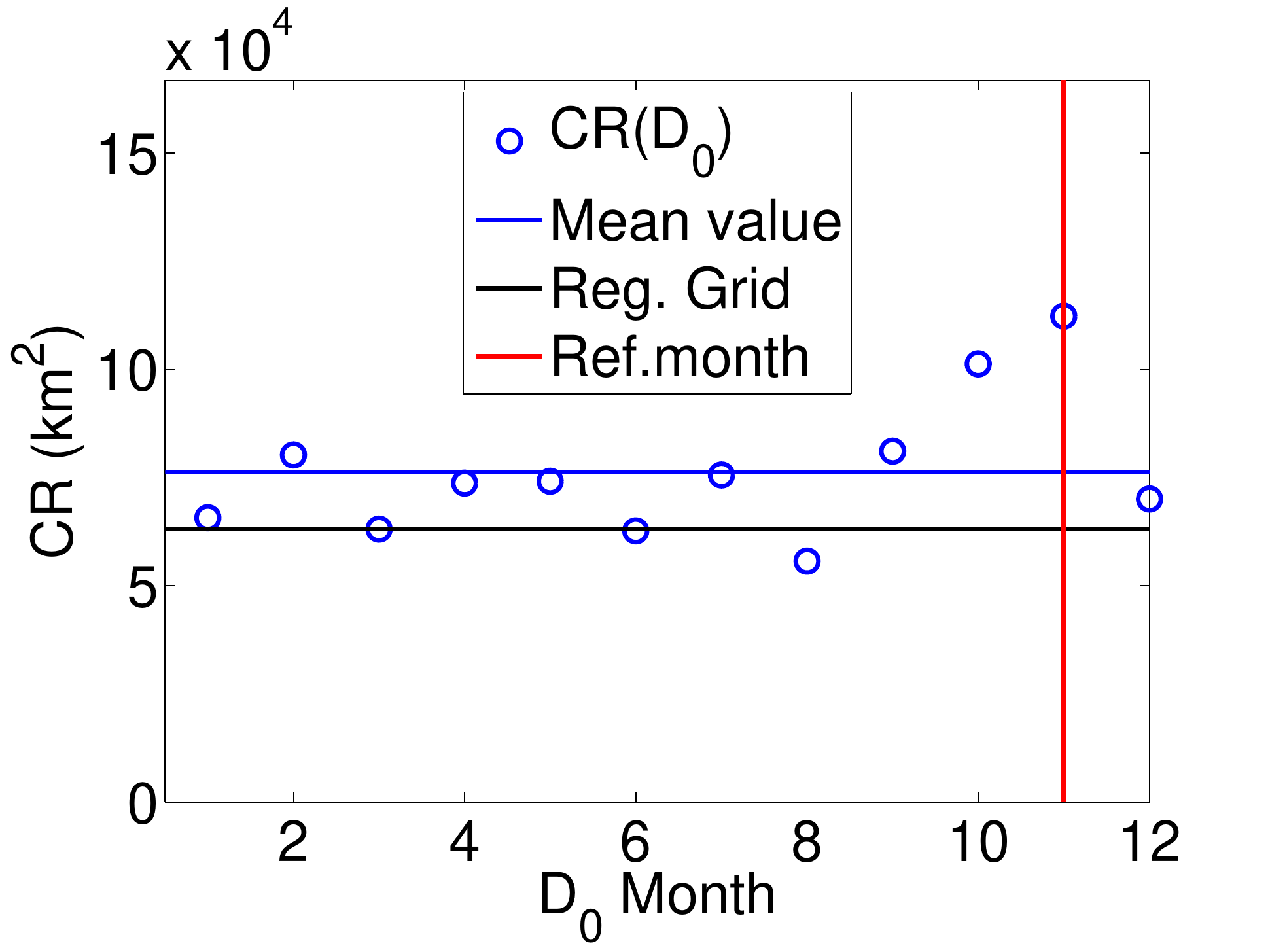}
\quad
\includegraphics[width=5.5cm]{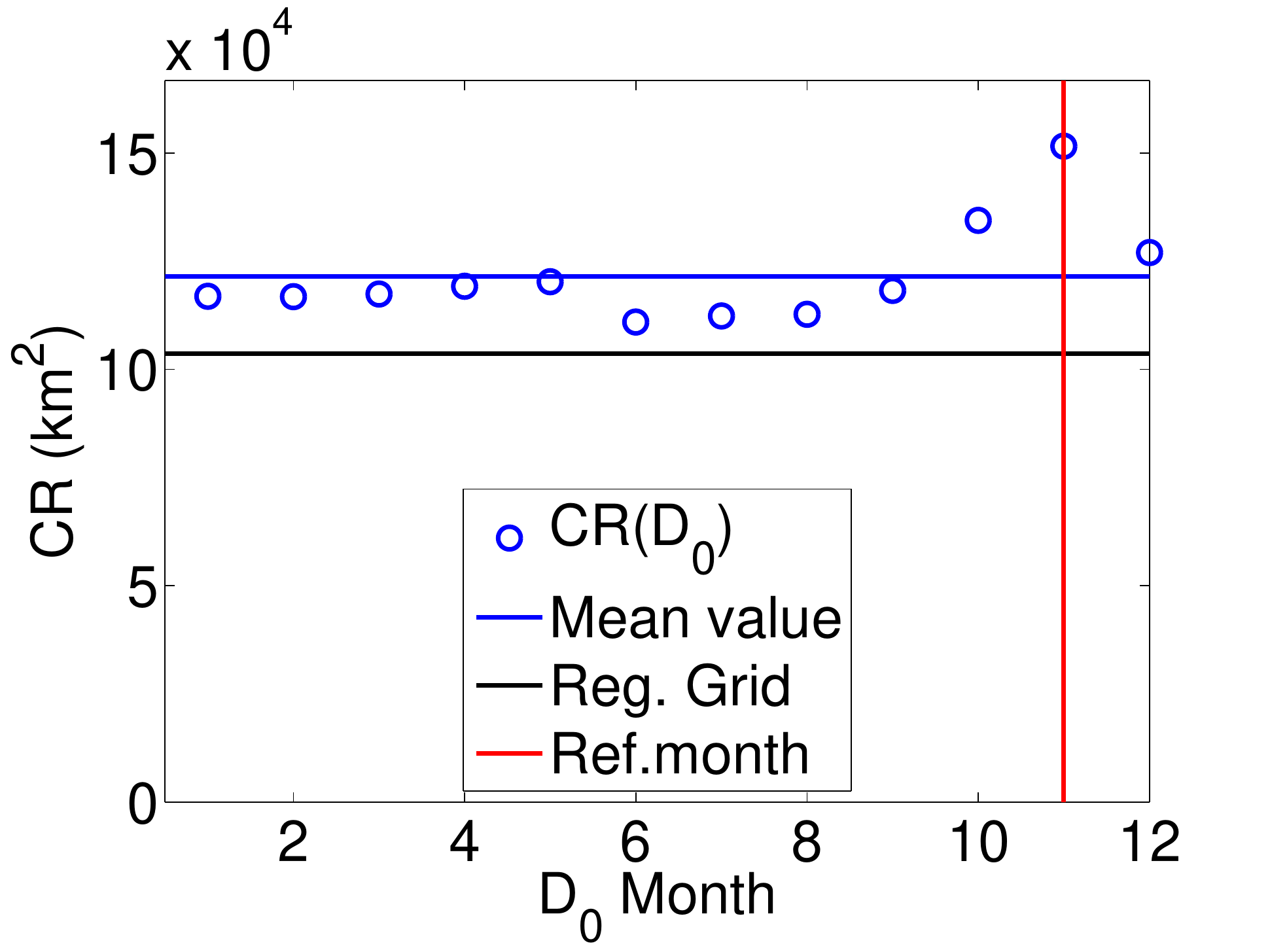}
\internallinenumbers
\caption{Surface of $\mathcal{R}_P$, advected from $D_R\,=\,$November, 11th, 2011,
monitored using the stations computed with velocity currents of different months along the year. $\delta_{IG}\,=\,\delta_{OG}\,=\,0.1\degree$, $\uptau=60\,days$. Left panel: $\sigma\,=\,0.1\degree$. Right panel: $\sigma\,=\,0.2\degree$
Horizontal lines: surface of November monitored using a
regular grid (black line) or 6 CR stations (blue line, mean value of the blue circles)}
\label{fig:an35NOturb}
\end{figure}

\textit{Surface monitored with stations computed along the
year.}

As in the former case, the region $\mathcal{R}_P$ is advected
starting from $D_0$ January, 11th, 2011,  for a period $\uptau$
of 30 and 60 days, and 6 CR stations were computed. This time
the stations are not used to see how many SVP drifters they
would have collected, but how many trajectories of the set
$\mathcal{T}(\mathcal{R})_{D_R\rightarrow D_R+\uptau}$, with
$D_R\,=\,$November, 11th, 2011, they would have intercepted.
$D_0$ is varied taking each time the 11th day of a different
month of 2011.

The results are reported in Fig. \ref{fig:an35NOturb} and
display in both cases a temporal decrease of the surface
sampled with the CR stations for the first three months.
Generally, the surface monitored with this method is
about $15\%$ greater than with a regular grid.

\section*{Appendix B. Relation with absolute velocity and mean kinetic
energy}
\label{sec:AppB}

A validation of Eqs. (\ref{eq_CR_v}) and (\ref{eq_CR_EK_2}) is
reported in Fig. \ref{fig:CR&EK&Uabs_mean}, where a map of
crossroadness, of kinetic energy and of absolute velocity
field, averaged for the month of November, is shown.  There,
the patterns look pretty identical. This is confirmed by the
scatter plot of Fig. \ref{fig:scatter_CRvsEKvsUabs}, in which
the expressions (\ref{eq_CR_v}) and (\ref{eq_CR_EK_2}) are
represented by the red line. The correlation coefficients of
the two plots are very similar, confirming the validation of
the assumptions leading to these equations.

Nevertheless, the kinetic energy or speed presents two main
differences with the crossroadness, that make them less
suitable for monitoring purposes in dispersion problems. In
fact, even if these quantities are very similar to CR when the
advected area is larger than the domain of calculation, this
ceases to be true for smaller advected domains. This is seen in
Fig. \ref{fig:CR&EK&Uabs_mean}, lower right panel. There, the
crossroadness is computed with the same parameters as in the
left upper panel ($\uptau=30$ days,
$\delta_{IG}\,=\,\delta_{OG}\,=\,\sigma=0.1\degree$) and on the same
\textit{observational grid}, but the \textit{initialization
grid} is the one used Subsec. \ref{subsec:svp}, i.e. much
smaller than the observational one. We can notice that the two
patterns differ radically, confirming the importance of the
fulfillment of the hypothesis leading to Eqs. (\ref{eq_CR_v}) and
(\ref{eq_CR_EK_2}). Furthermore, simple maps of kinetic energy or
speed do not allow to track the origin of the particles that
passed through each point, and thus to establish a hierarchy of
importance for observing stations.

\renewcommand{\theequation}{B.\arabic{equation}}
\setcounter{equation}{0}  
\renewcommand{\thefigure}{B.\arabic{figure}}
\setcounter{figure}{0}  
\renewcommand{\thetable}{B.\arabic{table}}
\setcounter{table}{0}  

\begin{figure}[h!]
\centering
\includegraphics[width=5.5cm]{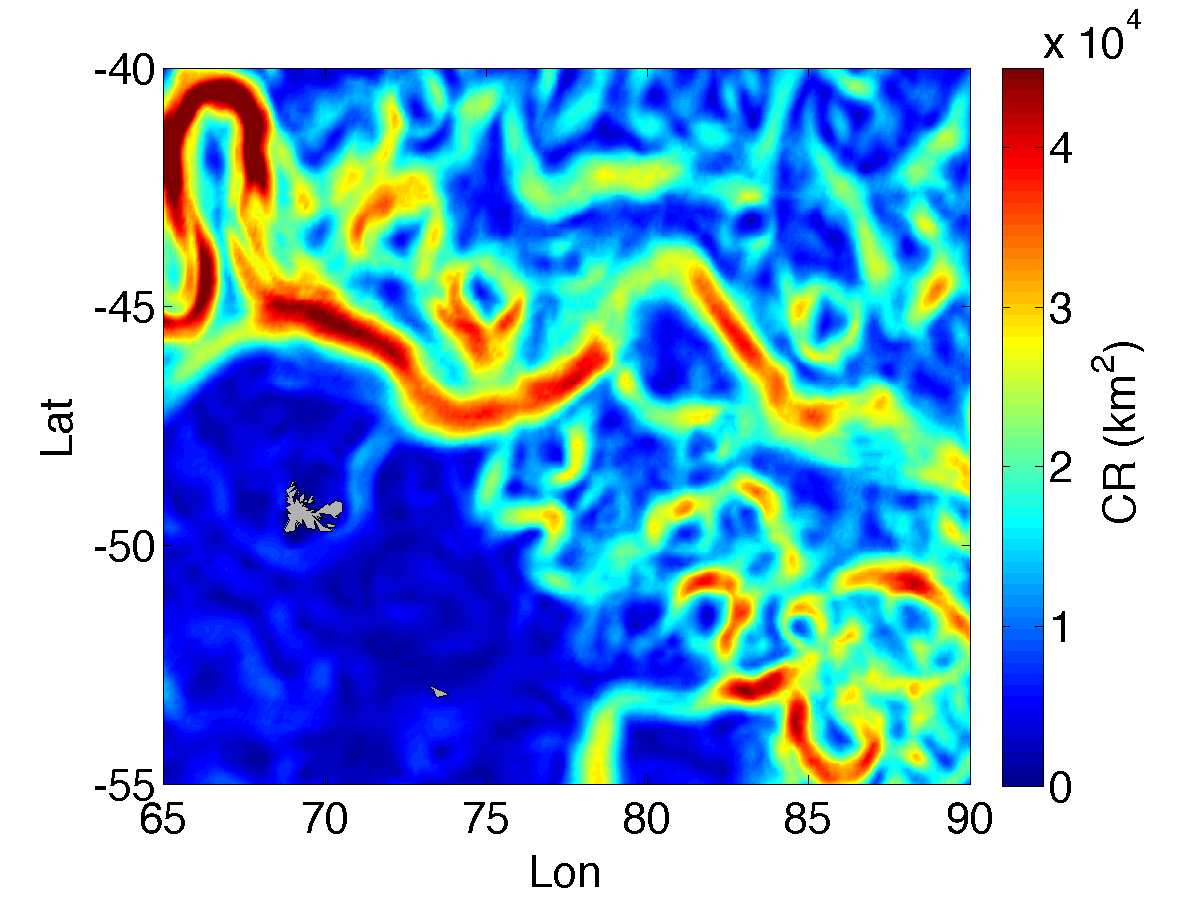}
\quad
\includegraphics[width=5.5cm]{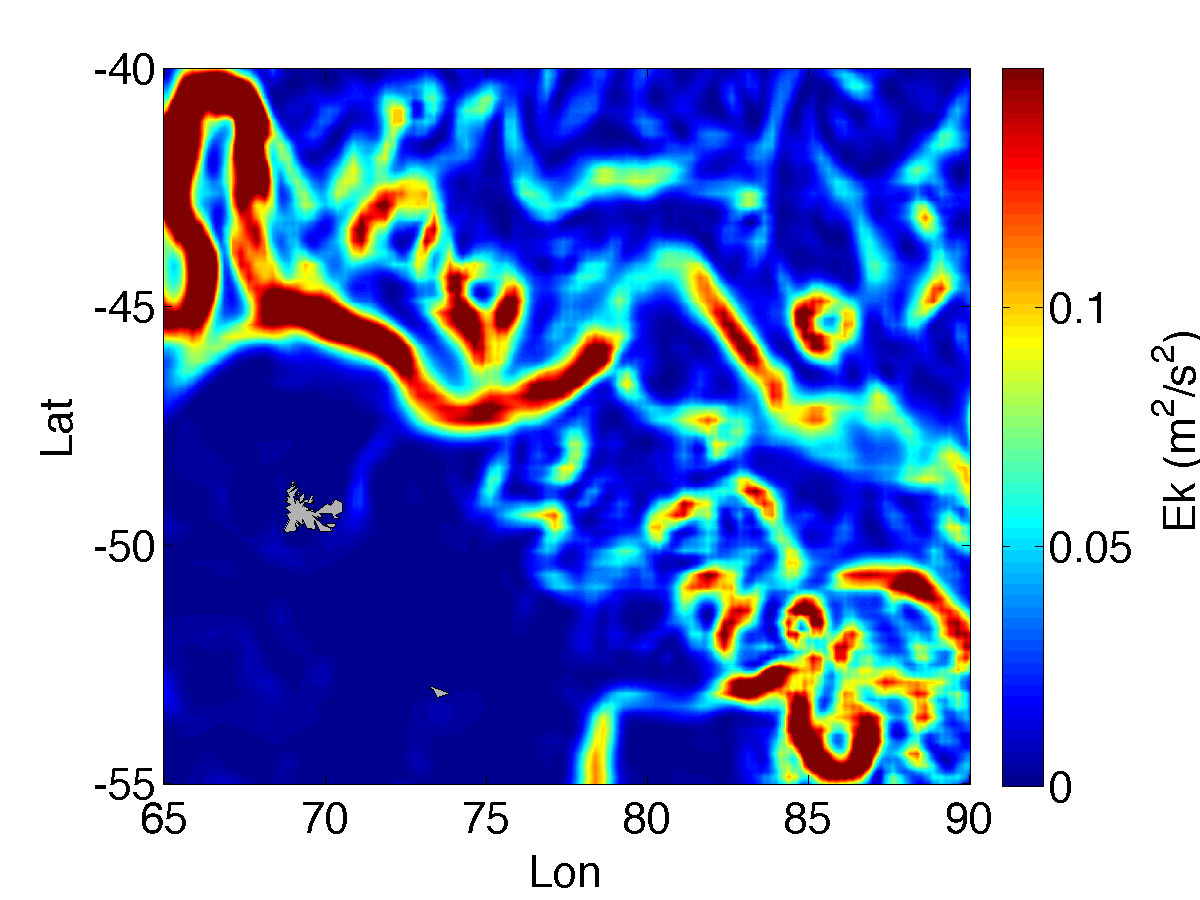}
\quad
\includegraphics[width=5.5cm]{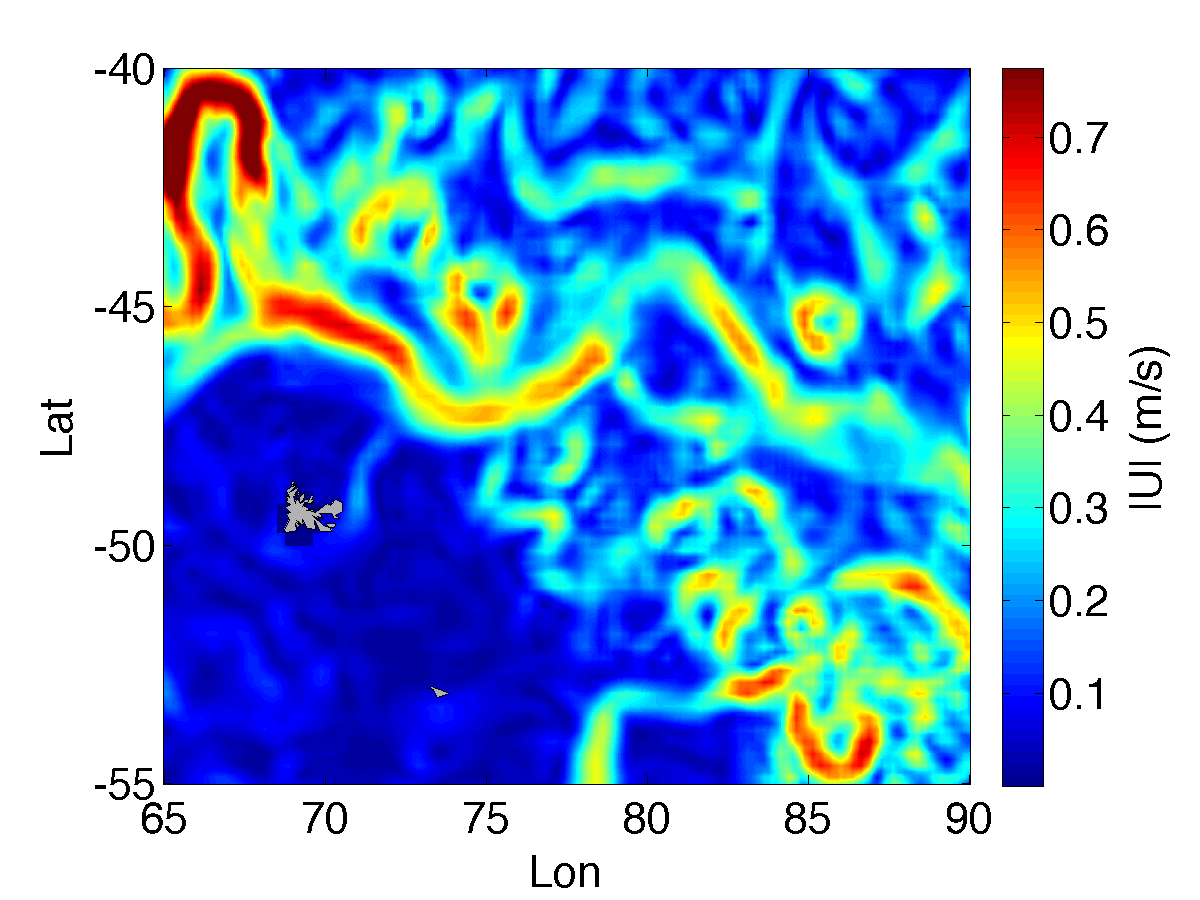}
\quad
\includegraphics[width=5.5cm]{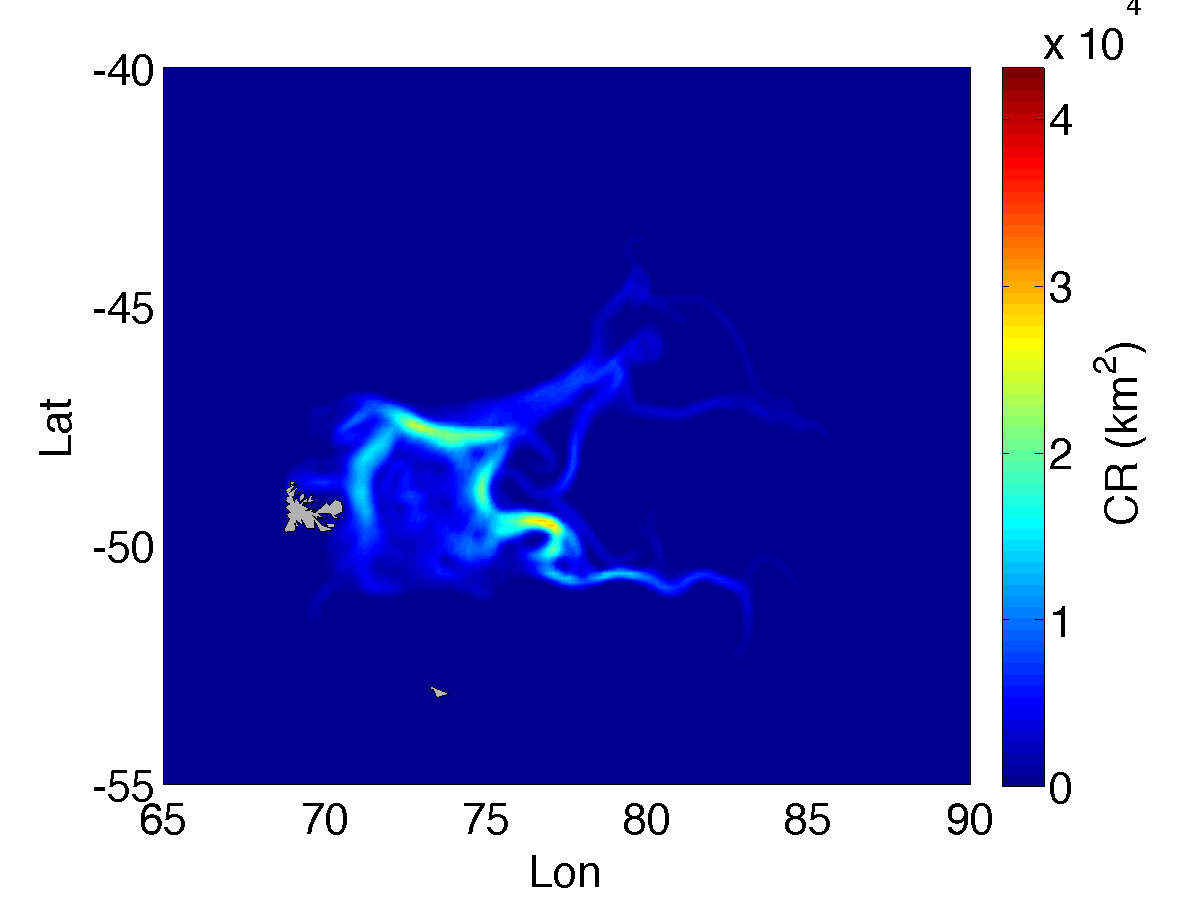}
\internallinenumbers
\caption{Top left panel: crossroadness relative to November 2011 computed
with an advection time $\uptau=30$ days, $\delta_{IG}\,=\,\delta_{OG}\,=\,\sigma=0.1\degree$.
Top right panel: mean kinetic energy $<E_K>$ of November 2011.
Lower left panel: mean absolute velocity field $<|v|>$ of November 2011.
Lower right panel: CR field computed over the same \textit{observational grid}
as in the left upper panel, with same parameters, but the \textit{initialization grid}
is only the drifter release region (\emph{rectangle} of longitude: [70, 75]$\degree$ and latitude:
[-51, -47]$\degree$) as in Subsec. \ref{subsec:svp}}
\label{fig:CR&EK&Uabs_mean}
\end{figure}

\begin{figure}[h]
\centering
\includegraphics[width=5.5cm]{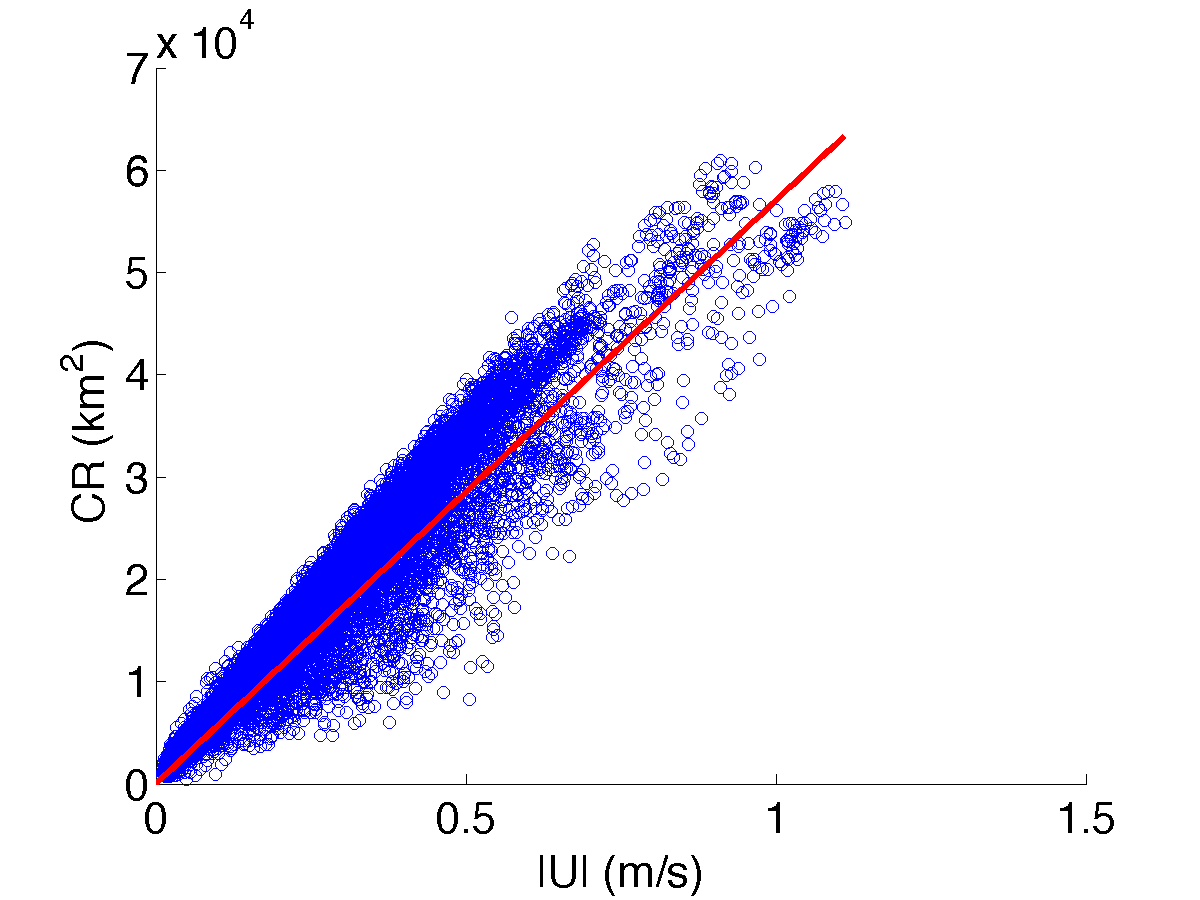}
\quad
\includegraphics[width=5.5cm]{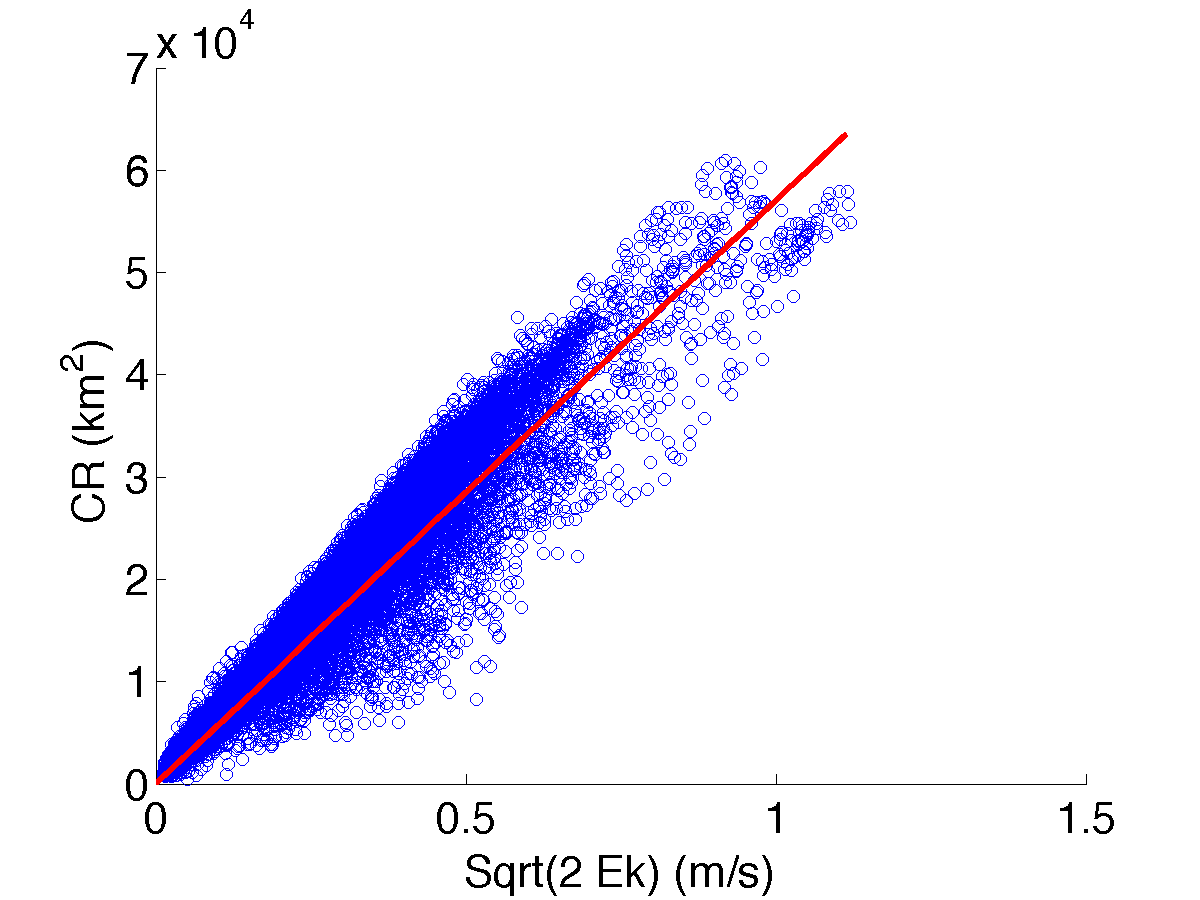}
\internallinenumbers
\caption{Scatterplot of values of CR vs $<|v|>$ (left panel, correlation coefficient: $0.969$)
and CR vs $\sqrt{2<E_K>}$ (right panel, correlation coefficient: $0.967$) of Fig. \ref{fig:CR&EK&Uabs_mean}
}
\label{fig:scatter_CRvsEKvsUabs}
\end{figure}

\label{Bibliography}
\bibliographystyle{deepsearesearchI}
\bibliography{Bibliography}  

\end{document}